\documentclass[a4paper,12pt]{iopart}
\usepackage{iopams}
\usepackage{graphicx}
\usepackage{epsfig}
\usepackage{amsfonts}
\usepackage{doublespace}
\usepackage{pennames}
\usepackage{sublabel}

\newcommand{\AmS}{{\protect\the\textfont2
  A\kern-.1667em\lower.5ex\hbox{M}\kern-.125emS}}
\newcommand{\decayarrow}{\makebox[0mm][l]{\rule{0.33em}{0mm}\rule[0.55ex]
 {0.044em}{1.55ex}}\rightarrow}

\newcommand{\mT}{\ensuremath{m_{\mathrm T}}}
\newcommand{\pT}{\ensuremath{p_{\mathrm T}}}

\newcommand{\dder}{\ensuremath{{\mathrm d}}}

\begin{document}
\begin{spacing}{1}

\title[Enhancement of hyperon production]{Enhancement of hyperon production 
at central rapidity
\mbox{in~158~$A$~GeV/$c$} Pb--Pb collisions}

\author{  
F~Antinori$^{l}$,
P~Bacon$^{e}$,
A~Badal{\`a}$^{f}$,
R~Barbera$^{f}$,
A~Belogianni$^{a}$,
W~Beusch$^{g}$,
I~J~Bloodworth$^{e}$,
M~Bombara$^{h}$,
G~E~Bruno$^{b}$,
S~A~Bull$^{e}$,
R~Caliandro$^{b}$,
M~Campbell$^{g}$,
W~Carena$^{g}$,
N~Carrer$^{g}$,
R~F~Clarke$^{e}$,
A~Dainese$^{l}$,
D~Di~Bari$^{b}$,
S~Di~Liberto$^{o}$,
R~Divi\`a$^{g}$,
D~Elia$^{b}$,
D~Evans$^{e}$,
G~A~Feofilov$^{q}$,
R~A~Fini$^{b}$,
P~Ganoti$^{a}$,
B~Ghidini$^{b}$,
G~Grella$^{p}$,
H~Helstrup$^{d}$,
K~F~Hetland$^{d}$,
A~K~Holme$^{k}$,
D~Huss$^{j}$,
A~Jacholkowski$^{f}$,
G~T~Jones$^{e}$,
P~Jovanovic$^{e}$,
A~Jusko$^{e}$,
R~Kamermans$^{s}$,
J~B~Kinson$^{e}$,
K~Knudson$^{g}$,
V~Kondratiev$^{q}$,
I~Kr\'alik$^{h}$,
A~Krav\v c\'akov\'a$^{i}$,
P~Kuijer$^{s}$,
V~Lenti$^{b}$,
R~Lietava$^{e}$,
R~A~Loconsole$^{b}$,
G~L\o vh\o iden$^{k}$,
V~Manzari$^{b}$,
M~A~Mazzoni$^{o}$,
F~Meddi$^{o}$,
M~E~Michalon-Mentzer$^{r}$,
A~Michalon$^{r}$,
M~Morando$^{l}$,
P~I~Norman$^{e}$,
A~Palmeri$^{f}$,
G~S~Pappalardo$^{f}$,
B~Pastir\v c\'ak$^{h}$,
R~J~Platt$^{e}$,
E~Quercigh$^{l}$,
F~Riggi$^{f}$,
D~R\"ohrich$^{c}$,
G~Romano$^{p}$,
K~\v{S}afa\v{r}\'{\i}k$^{g}$,
L~\v S\'andor$^{h}$,
E~Schillings$^{s}$,
G~Segato$^{l}$,
M~Sen\'e$^{m}$,
R~Sen\'e$^{m}$,
W~Snoeys$^{g}$,
F~Soramel$^{l}$
\footnote[2]{Permanent
address: University of Udine, Udine, Italy},
M~Spyropoulou-Stassinaki$^{a}$,
P~Staroba$^{n}$,
M~Thompson$^{e}$,
R~Turrisi$^{l}$,
T~S~Tveter$^{k}$,
J~Urb\'{a}n$^{i}$,
P~van~de~Ven$^{s}$,
P~Vande~Vyvre$^{g}$,
A~Vascotto$^{g}$,
T~Vik$^{k}$,
O~Villalobos~Baillie$^{e}$,
L~Vinogradov$^{q}$,
T~Virgili$^{p}$,
M~F~Votruba$^{e}$,
J~Vrl\'{a}kov\'{a}$^{i}$\ and
P~Z\'{a}vada$^{n}$
}
\address{
$^{a}$ Physics Department, University of Athens, Athens, Greece\\
$^{b}$ Dipartimento IA di Fisica dell'Universit{\`a}
       e del Politecnico di Bari and INFN, Bari, Italy \\
$^{c}$ Fysisk Institutt, Universitetet i Bergen, Bergen, Norway\\
$^{d}$ H{\o}gskolen i Bergen, Bergen, Norway\\
$^{e}$ University of Birmingham, Birmingham, UK\\
$^{f}$ University of Catania and INFN, Catania, Italy\\
$^{g}$ CERN, European Laboratory for Particle Physics, Geneva, Switzerland\\
$^{h}$ Institute of Experimental Physics, Slovak Academy of Science,
              Ko\v{s}ice, Slovakia\\
$^{i}$ P J \v{S}af\'{a}rik University, Ko\v{s}ice, Slovakia\\
$^{j}$ GRPHE, Universit\'{e} de Haute Alsace, Mulhouse, France \\
$^{k}$ Fysisk Institutt, Universitetet i Oslo, Oslo, Norway\\
$^{l}$ University of Padua and INFN, Padua, Italy\\
$^{m}$ Coll\`ege de France, Paris, France\\
$^{n}$ Institute of Physics, Prague, Czech Republic\\
$^{o}$ University ``La Sapienza'' and INFN, Rome, Italy\\
$^{p}$ Dipartimento di Scienze Fisiche ``E.R. Caianiello''
       dell'Universit{\`a} and INFN, Salerno, Italy\\
$^{q}$ State University of St. Petersburg, St. Petersburg, Russia\\
$^{r}$ IReS/ULP, Strasbourg, France\\
$^{s}$ Utrecht University and NIKHEF, Utrecht, The Netherlands
}
\ead{Giuseppe.Bruno@ba.infn.it, Bruno.Ghidini@ba.infn.it}
\begin{abstract}

Results are presented on hyperon and antihyperon production in Pb--Pb, pPb and pBe collisions at 158 GeV/c per nucleon. 
$\Lambda$, $\Xi$ and $\Omega$ yields have been measured at central rapidity and medium transverse momentum as functions of the centrality of the collision.
Comparing the yields in Pb--Pb to those in pBe interactions, strangeness enhancement is observed. The 
enhancement increases with the centrality and with the strangeness content of the hyperons, reaching a factor of about 20 for the $\Omega$ in the central Pb--Pb collisions.
\end{abstract}



\vspace{-0.4cm}
\section{Introduction}
Collisions between heavy nuclei at very high energy provide a unique opportunity 
of producing and studying in laboratory the QCD-predicted phase transition from ordinary 
nuclear matter to a colour-deconfined 
Quark-Gluon Plasma (QGP)~\cite{cabibbo-QM2004}.   
The plasma occurrence and its properties can only be inferred from a study of the characteristics 
of the colourless 
particles 
finally emerging from the collision.  
Several experimental signatures for the QGP have been proposed, among which  
the production of strange particles, in particular, is considered to be a powerful diagnostic tool~\cite{rafelski}.  
If a QGP state is formed, an increased production of ${\rm s}$\ and ${\rm \bar{s}}$\ quarks with respect to  
normal hadronic interactions is expected, because the production of ${\rm  s \bar{s}}$\  pairs
becomes competitive   
with that of ${\rm u \bar{u}}$\ and ${\rm d \bar{d}}$\  pairs due to partial chiral symmetry restoration 
and, at SPS energies, also because of the Pauli blocking 
of the production of light quarks  
in a ${\rm  d }$\ and ${\rm u }$\ rich  environment. 
The production of ${\rm s \bar{s}}$\ pairs 
is expected to equilibrate 
in a few fm/$c$, comparable with the plasma lifetime. The net result,   
after statistical hadronisation, would be an enhancement of strange and multistrange  
particle production in ion--ion interactions with respect to nucleon-nucleon interactions. 

An increased production of strange particles would also be possible in a purely hadronic 
scenario (hot and dense hadron gas) where the relative abundance of strange particles 
could grow gradually in a chain of rescattering processes. Such a mechanism is however 
hindered by the high threshold for associated production and, therefore,  
the characteristic time scale would be much longer than in a QGP, 
particularly  
for multistrange antibaryons.  

The NA57 experiment~\cite{proposal} at the CERN SPS has been designed to study the 
strange baryon enhancement in Pb--Pb collisions already observed by its 
predecessor WA97\cite{WA97proposal} with the hierarchy expected in a QGP scenario: 
${\mathrm \Omega}$ being more enhanced than ${\mathrm \Xi}$ and ${\mathrm \Xi}$ more than \PgL.
 In particular, the aim of NA57 is to   
investigate the dependence of the strangeness enhancement on the collision 
centrality (measured by the number of nucleons taking part in the collision) and also on the collision 
energy. For these reasons the NA57 experiment has extended the centrality range covered by 
WA97 towards lower centrality and has collected Pb--Pb data at two beam momenta, 158 and 
40 $A$\ GeV/$c$. As a reference  for the enhancement evaluation, we use hyperon production 
in pBe interactions, which provide a good approximation to elementary nucleon-nucleon interactions. 
NA57 has collected data on pBe interactions at 40 GeV/$c$\ only, while at 158 GeV/$c$\ the pBe 
data available from the WA97 experiment have been used. In this paper, results on strange baryon 
and antibaryon production at the upper energy are presented as function of the centrality. 
In addition, hyperon production in pPb interactions at 158 $A$\ GeV/$c$\ 
(from the WA97 experiment) is presented for comparison.  

The paper is organised as follows. The NA57 apparatus is described in section 2, 
along with 
that for WA97, as used for the pBe and pPb data taking. 
In section 3 the strange particle 
reconstruction procedure is explained and the extraction of clean samples of the hyperons under 
study is discussed. Section 4 deals with the evaluation of yields and enhancements as 
functions of the centrality. A discussion of the results 
is presented in section 5. Finally, conclusions are drawn in section 6.  
\section{The experiments}
\begin{figure}[p]
\centering
\resizebox{0.97\textwidth}{!}{
\includegraphics{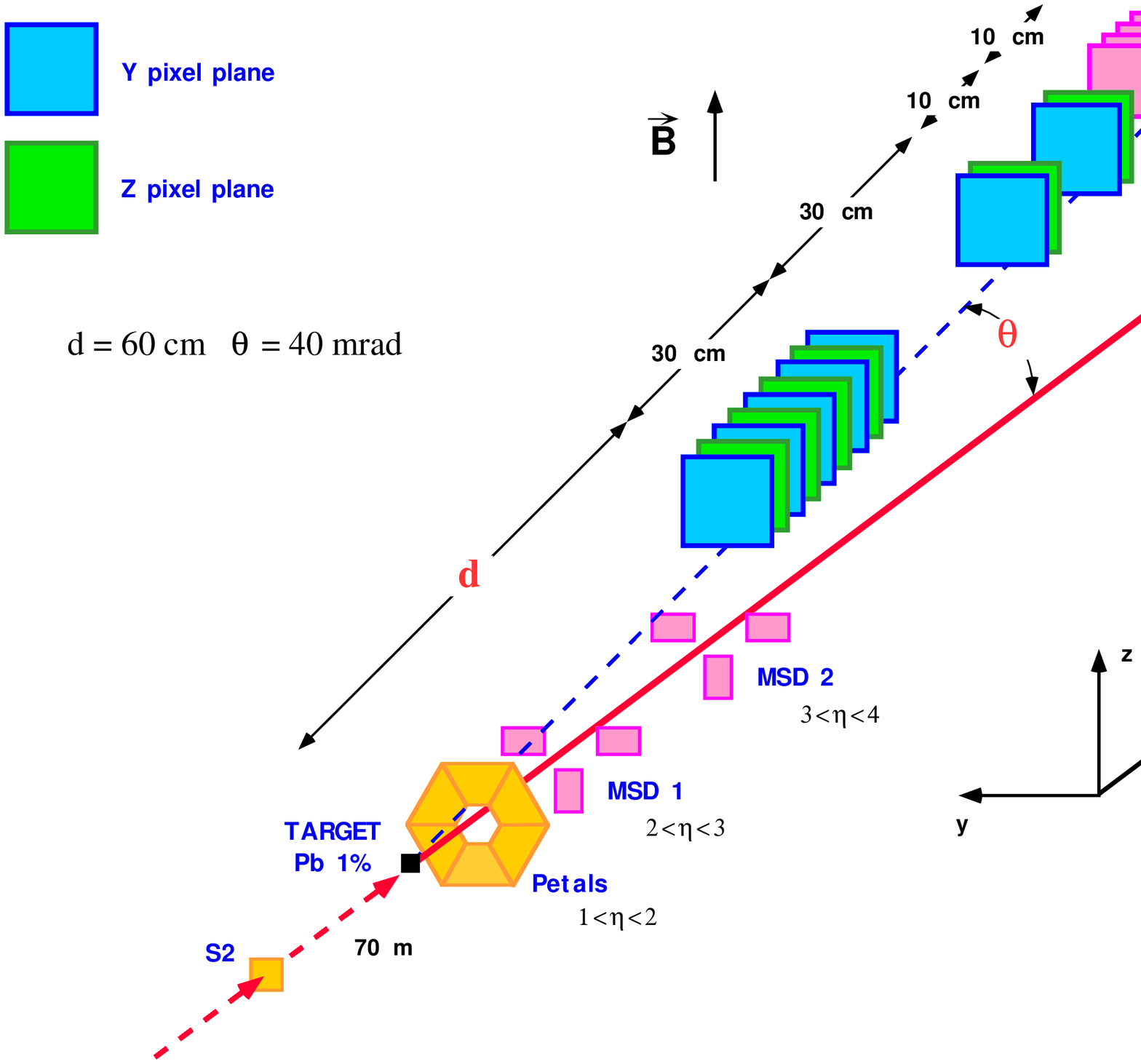}}
\caption{The NA57 apparatus placed inside the 1.4~T field of the GOLIATH
 magnet. 
\label{1a}}
\vspace{0.4cm}
\centering
\resizebox{0.97\textwidth}{!}{
\includegraphics{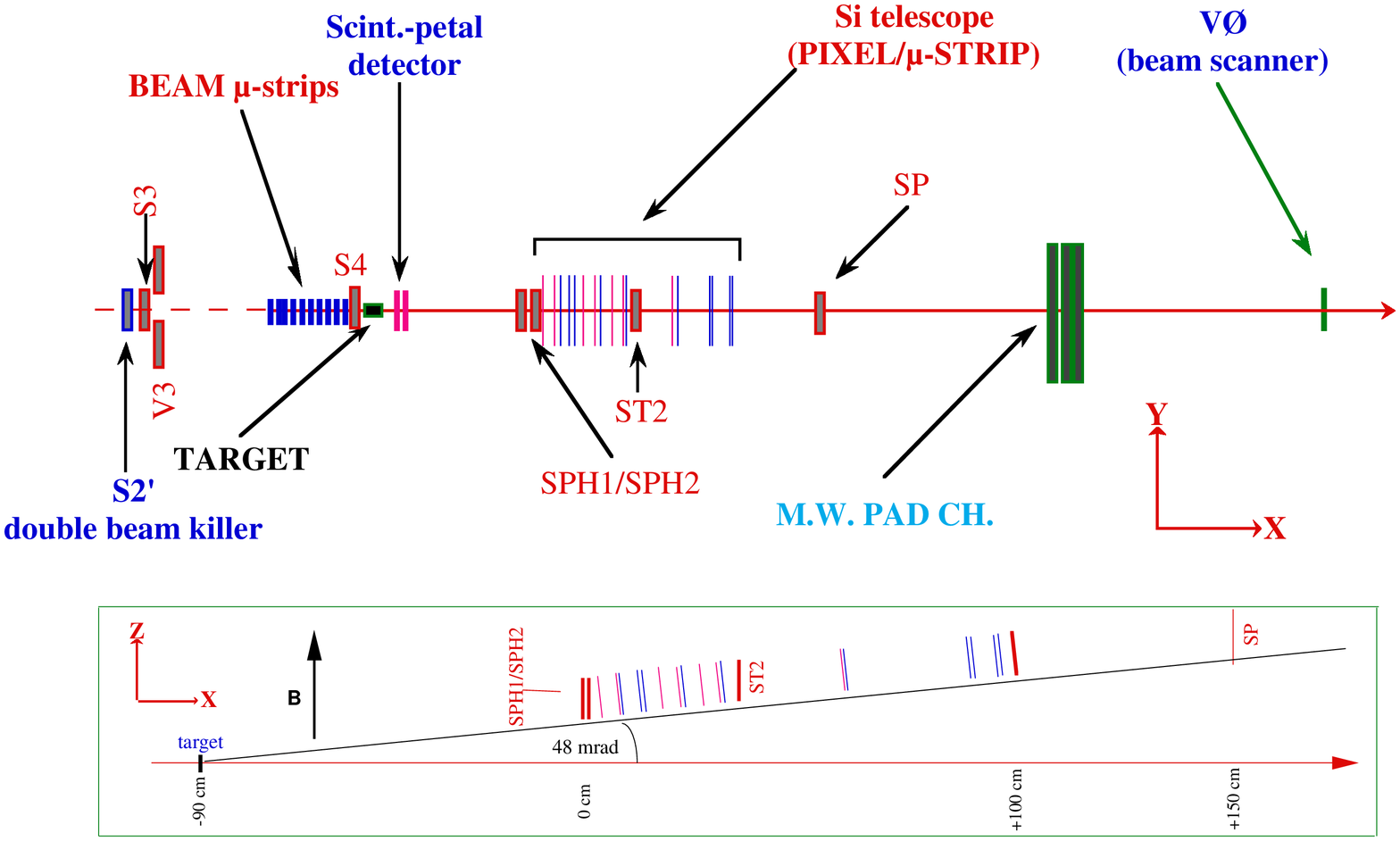}}
\caption{The WA97 apparatus placed inside the 1.8~T field of the OMEGA magnet.
\label{1b}}
\end{figure}

The layouts for the NA57 and WA97 experiments (figures~\ref{1a},\ref{1b}) are 
conceptually similar~\cite{na57exp,wa97exp}.  
In both experiments the charged tracks coming from strange particle decays are reconstructed 
in a telescope made from an array of silicon detector planes of 5x5 cm$^2$\ cross-section 
placed in an approximately uniform magnetic field perpendicular to the beam line; the bulk of the detectors was   
closely packed in approximately 30 cm length, this compact part 
being used for pattern recognition. 
The telescope was placed above the beam line, inclined and aligned with the lower edge of the 
detectors laying on a line pointing to the target.
The inclination angle $\theta$\ and the distance $d$\ of the first plane 
from the target are set so as to accept 
particles produced 
in about a unit of rapidity around mid-rapidity, with 
transverse momentum above a few hundred MeV/$c$.  
In such conditions the track densities 
reach about $10 \;{\rm cm}^{-2}$\ for central Pb--Pb collisions at 158 GeV/$c$. The 
detectors employed have good 
granularity and precision, allowing them to handle such high particle densities.  
The two experimental setups are outlined in the following.  
\sublabon{figure}
\sublaboff{figure}

\begin{figure}[b]
\centering
\huge{a)}
\resizebox{0.40\textwidth}{!}{
\includegraphics{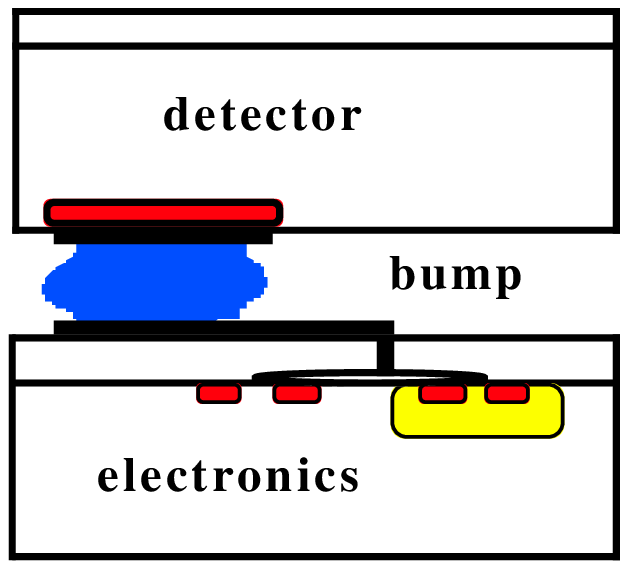}} \hspace{1.0cm}
\huge{b)}
\resizebox{0.38\textwidth}{!}{
\includegraphics{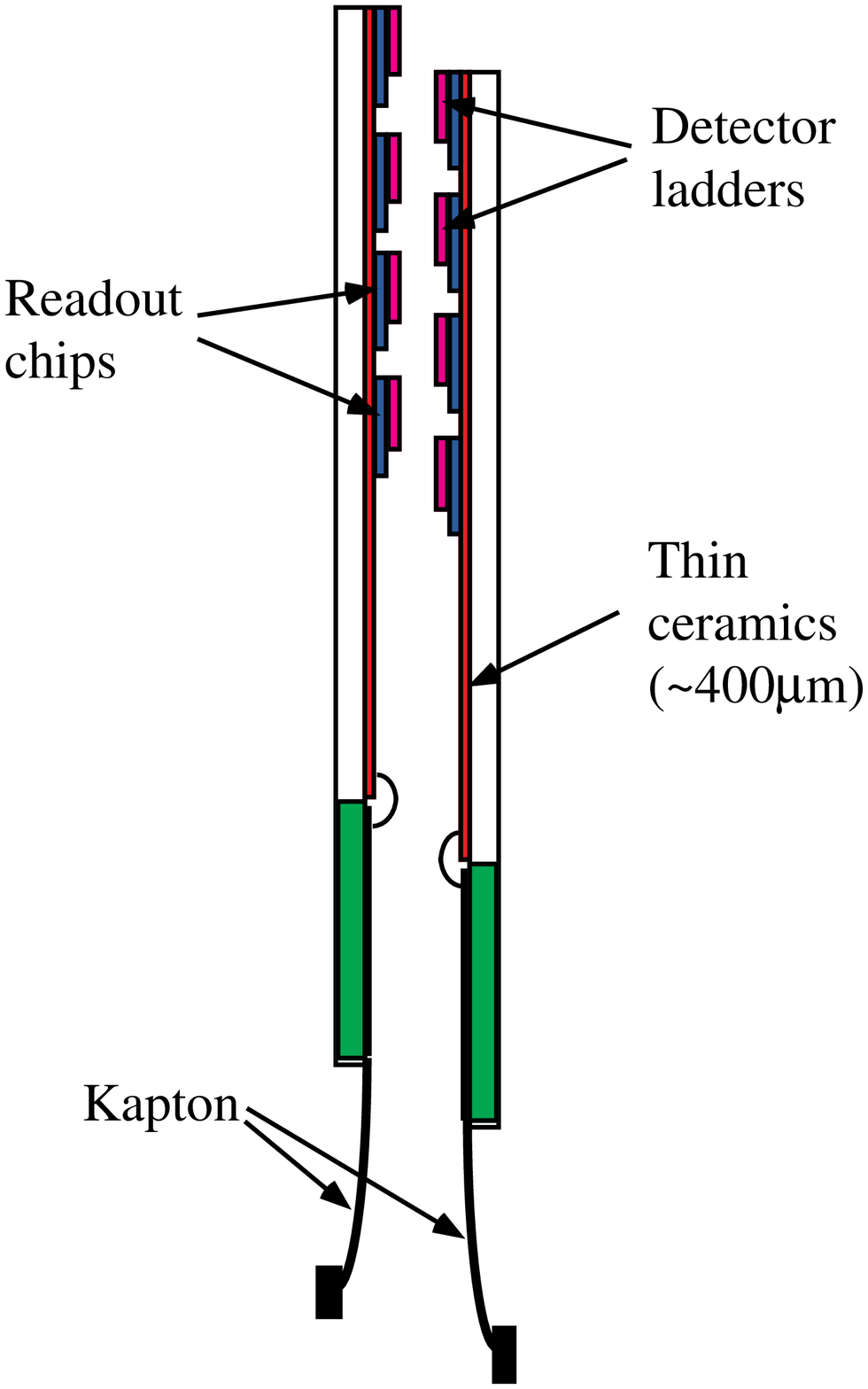}}
\caption{Flip-chip assembly. 
\label{2a}}
\caption{Two arrays mounted face to face and staggered to form a plane.
\label{2b}}
\end{figure}

The 
NA57 telescope (figure~\ref{1a}) was entirely made of hybrid silicon pixel detectors. 
This technique was successfully pioneered by the WA97 experiment together 
with the RD19 collaboration~\cite{RD19}. 
A hybrid pixel detector is a matrix of reverse-biased silicon detector diodes, 
bump-bonded to front-end  electronics chips, 
as sketched in figure~\ref{2a}. 
Two generations of silicon pixel detectors 
were used: Omega2~\cite{Omega2}, with a pixel size of $75 \times 500 \, \mu m^2$, and 
Omega3~\cite{Omega3}, with a pixel size of $50 \times 500 \, \mu m^2$. In both cases, six 
integrated front-end/readout chips were bump-bonded to a silicon sensor segmented into pixels, 
to form the basic building block of the detector, called a ``ladder''. Several ladders, six for the 
Omega2 and four for the Omega3, were then glued on a ceramic carrier;  each plane of the 
telescope consisted of two such arrays mounted face-to-face, suitably staggered, to hermetically 
cover a sensitive area of $5 \times 5 \; {\rm cm}^2$\ with active elements, as shown in figure~\ref{2b}. 
In total, the telescope contained seven Omega2 and six Omega3 planes, 
with the long pixel edge alternately oriented along the magnetic field and the bend direction 
(z and y respectively in figure~\ref{1a}), for a total of about $10^6$\ channels. 
Nine of them were in the 30 cm long compact part of the telescope, while the remaining four pixel 
planes together with four double-sided silicon microstrip detectors 
(95 $\mu$m pitch, stereo angle $\pm 17.5^o$)  with analogue read-out formed a lever arm system, 
placed downstream in the magnetic field to improve the momentum resolution of the high momentum tracks. 
An array of 6 scintillation counters (Petals), located 10 cm downstream of the target, provided a fast signal to trigger on central collisions. The Petals covered the pseudorapidity region $1<\eta<2$\  
and their thresholds were set so as to accept events with track multiplicities above an adjustable limit. 
This was tuned so that the triggered event sample correspond to approximately the most central 
60\% of the Pb--Pb inelastic cross-section. 
The thickness of the Pb target was of 1\% of the interaction length, 
which made the trigger rate and the 
data acquisition capability properly matched.
The centrality of the collisions has been determined off-line by analysing the charged particle 
multiplicity measured by two stations of microstrip silicon detectors (MSD) which together sampled 
the pseudorapidity interval $2<\eta<4$. The whole apparatus was placed inside the 1.4~T magnetic 
field of the GOLIATH magnet in the CERN North experimental Area. A total sample of about 
$460\times 10^6$ Pb--Pb interactions at 158 $A$\ GeV/$c$\ has been recorded.  
\sublabon{figure}
\sublaboff{figure}

For the WA97 experiment the 1.8~T OMEGA magnet in the CERN West experimental 
Area was used. The telescope consisted of 10 silicon microstrip planes (50 $\mu$m pitch with 
binary readout) with the strips oriented alternately in the horizontal and vertical direction, 
interleaved with seven pixel planes providing altogether $0.5 \times 10^6$\ detecting elements 
(figure~\ref{1b}). The compact part consisted of six pixel and five microstrip planes, while the 
lever arm system used the remaining pixel and microstrip planes distributed along 
the telescope line inside the magnet, followed by three multi-wire proportional chambers 
with pad cathode readout located just outside the magnetic field. In the pBe and pPb runs,  
scintillation counters placed before and behind the compact part of the 
telescope have been used in the trigger to require at least one track traversing the telescope; 
for most of the data taking, the additional condition of two tracks entering the telescope was 
applied in order to increase the recorded sample of 
hyperon events. The protons in the beam were 
selected using two \v{C}erenkov CEDAR detectors~\cite{cedar}. In order to allow the 
event-by-event evaluation of the position of the main interaction vertex, 
an array of 10 microstrip detectors of 20 $\mu$m pitch 
was mounted before the 8\% interaction length beryllium target with the strips oriented alternately 
in the horizontal and vertical direction. 
This beam telescope was not present for the pPb run; in this case, however, a shorter target was 
used and therefore the position of the primary vertex was better constrained  
due to the smaller thickness of the target.  
The total recorded data sample consists of $219\times 10^6$\ pBe and $287\times 10^6$\ 
pPb interactions at 158 GeV/$c$\ beam momentum. 
\section{Data analysis}
\subsection{Extraction of the signals}
The \PgL, \PgXm, \PgOm\ hyperons and their antiparticles were identified  
through their weak decays into charged particles only, namely: 
\begin{equation}
\label{eq:decay}
\begin{array}{lllll}
 \hspace{-1.4cm}
    \Lambda \; \rightarrow \;  {\rm p} + \Pgpm    & \hspace{10mm}& 
    \PgXm  \; \rightarrow  \;  \PgL + \pi^- &  \hspace{10mm}&  \Omega^- \; \rightarrow \;
   \Lambda + \PKm  \\
       &       &  
 \hspace{1.4cm} \decayarrow  {\rm p} + \Pgpm  &    & \; \hspace{1.3cm} \decayarrow  {\rm p} + \Pgpm   \\
\end{array}
\end{equation}
with the corresponding charge conjugates for the anti-hyperons; all the secondary 
charged tracks were required to be reconstructed in the compact part of the telescope. 
The physical signals have been extracted using geometrical and kinematical constraints 
with a procedure similar to that explained in reference~\cite{selection} and described below.  
Slightly different selection criteria 
have been employed for Pb--Pb, pBe and pPb analysis.
In the following we discuss the most important ones.  
The complete sets of conditions and numerical values of the cuts for 
both 
\PgL\  
and cascade selection are described in reference~\cite{giuseppe_tesi} for Pb--Pb,
in~\cite{norman_tesi} for pBe, in~\cite{rocco_tesi} for pPb.  

The \PgL\  and \PagL\ hyperons\footnote{The apparatus being not able to detect the 
electromagnetic decay $\Sigma^0 \rightarrow \Lambda \gamma$, in the following 
the symbol ``$\Lambda$'' indicates the unresolved $\Lambda / \Sigma^0$\ signal.} 
(as well as the \PKzS\ mesons) have been searched for by combining any pair of 
oppositely charged tracks in an event to find what is usually called a ${\rm V^0}$\ 
topology: 
the two tracks were required to cross in space within a predetermined tolerance 
(distance of closest approach smaller than 0.3 mm in the Pb--Pb sample, 
1 mm in pBe and pPb), in a fiducial 
region located between the target and the telescope, 
whose actual dimensions were tuned separately for each 
experimental setup (e.g., 30 cm length in Pb--Pb).  
The ${\rm V^0}$\ line of flight was required to point to the vertex 
within a transverse region which extends, e.g. for Pb--Pb, 
to $\pm 0.9$\ mm in $y$\ and $\pm 1.8$\ mm in $z$.  
To improve the signal quality,    
only those ${\rm V^0}$s have been retained for which the tracks cross 
again in the bending-plane projection after leaving the decay vertex 
(``cowboy'' topology).  
The \PgL\ (\PagL) candidates were required to lie in the regions of the Podolanski-Armenteros  
plot~\cite{armenteros}   $|\alpha| >0.45$\footnote[5]{The $\alpha$\ variable is defined as 
$\alpha=(q_L^+ - q_L^-)/(q_L^+ + q_L^-)$ \ where $q_L^+$\ and $q_L^-$\ represent respectively 
the positive and negative decay track momentum component along the direction of motion 
of the ${\rm V^0}$. A detailed description of the use of the 
Podolanski-Armenteros variables to separate \PgL\ , \PagL\ and \PKzS\ can be found in 
references~\cite{PLBWA85,giuseppe_tesi}.}, 
eliminating the bulk of the contamination from \PKzS\  decays; 
the remaining contamination was then removed by demanding that the effective mass 
combination of the pair, treated as a $\pi^+$$\pi^-$\ pair, differs by more than $\simeq30$ 
MeV/$c^2$\ from the nominal \PKz\ mass. 
Finally, to remove background from photon conversions the transverse momentum of the decay 
tracks with respect to the ${\rm V^0}$\ line of flight was required to be greater than 
a few tens of MeV/$c$.  

\begin{figure}[h]
\centering
\resizebox{0.80\textwidth}{!}{%
\includegraphics{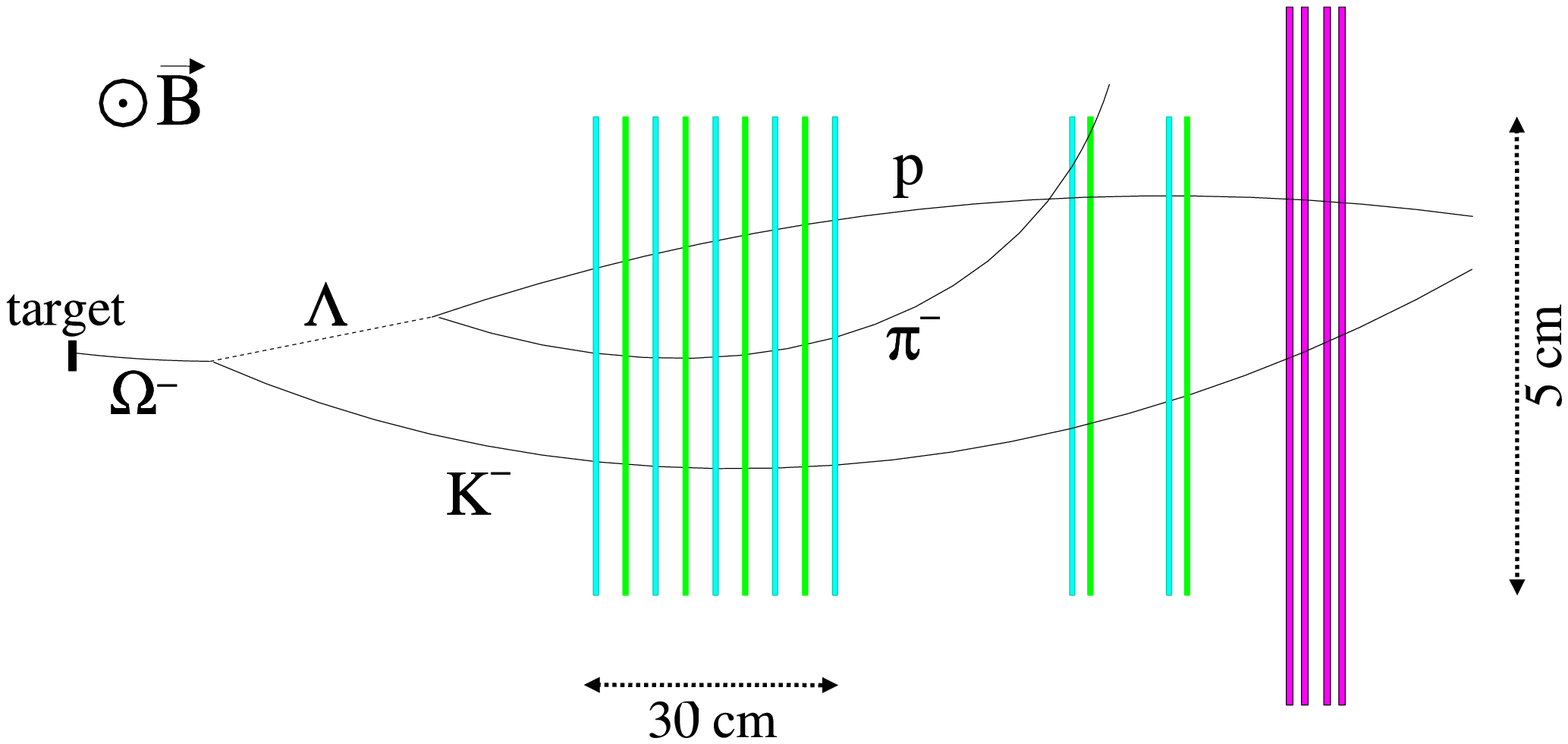}}
\caption{Example of the reconstruction of a cascade decay (not to scale).}
\label{2}
\end{figure}

The cascade candidates were selected by combining a ${\rm V^0}$\ (required not to point to the vertex) 
with a charged track of the proper sign and requiring a set of conditions similar to those employed 
for the ${\rm V^0}$s (e.g., on the distance of closest approach between the ${\rm V^0}$\ and the charged track, 
on the fiducial region, on the impact parameter at the vertex).   
In addition, the ${\rm V^0}$\ decay vertex was required to be located downstream of the
cascade decay vertex and the $\Omega$ candidates were required not to be kinematically 
ambiguous with the decay of a $\Xi$.  
Figure~\ref{2} shows as an example a sketch of the reconstruction of the cascade decay for an \PgOm\ .

Accurate knowledge of the position of the production vertex is crucial both for the selection 
analysis and for the subsequent correction procedure, explained in the next section. 
In the analysis of the pBe sample the vertex position has been calculated for 
each candidate by minimising the distance in space between the hyperon line of flight and the 
beam track measured by the beam telescope. In the Pb--Pb data sample, where the track 
multiplicity is large and a thinner target was used, the vertex has been determined on an 
event-by-event basis by averaging the intersections of the tracks extrapolated back to the target 
plane and then taking the average 
over a ``run'' (i. e., a data portion containing $\simeq10^4$ events). 
Finally, for the pPb data, where the track multiplicity is comparable to that for the pBe collisions but 
the longitudinal position of the vertex was better constrained due to the shorter target, 
the average over several events has been taken.
 
\begin{figure}[p]
\centering
\resizebox{0.82\textwidth}{!}{%
\includegraphics{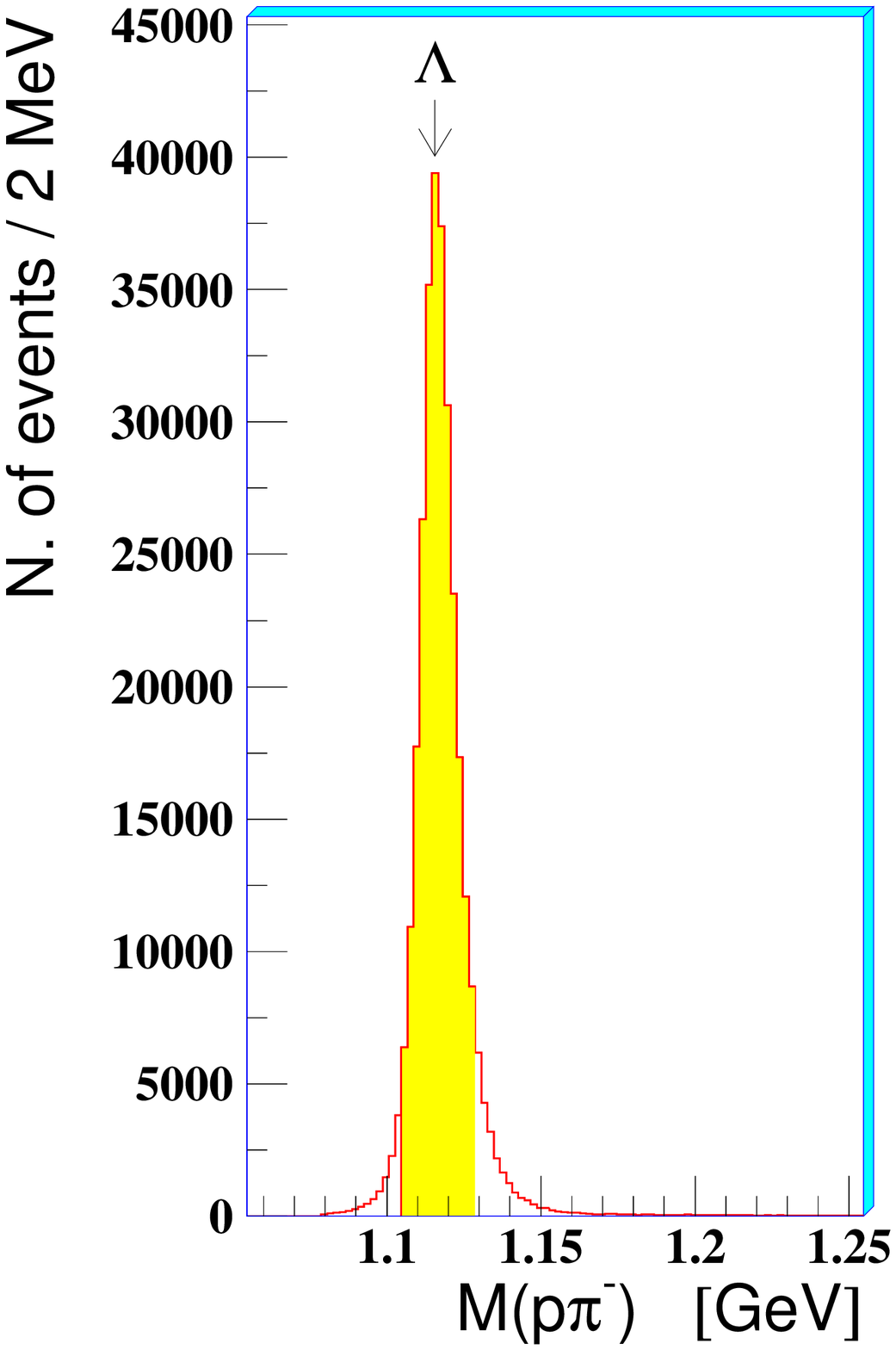}
\includegraphics{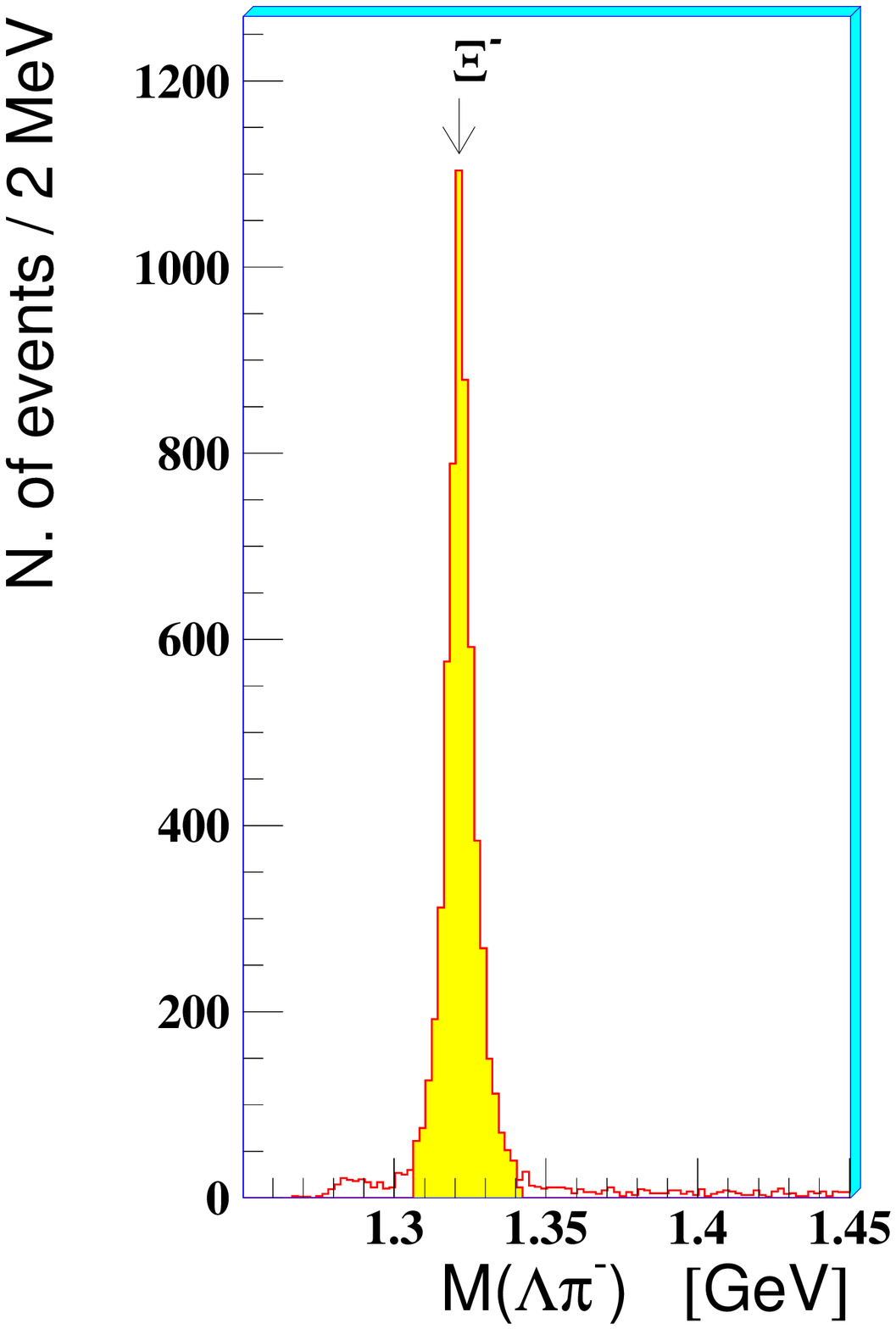}
\includegraphics{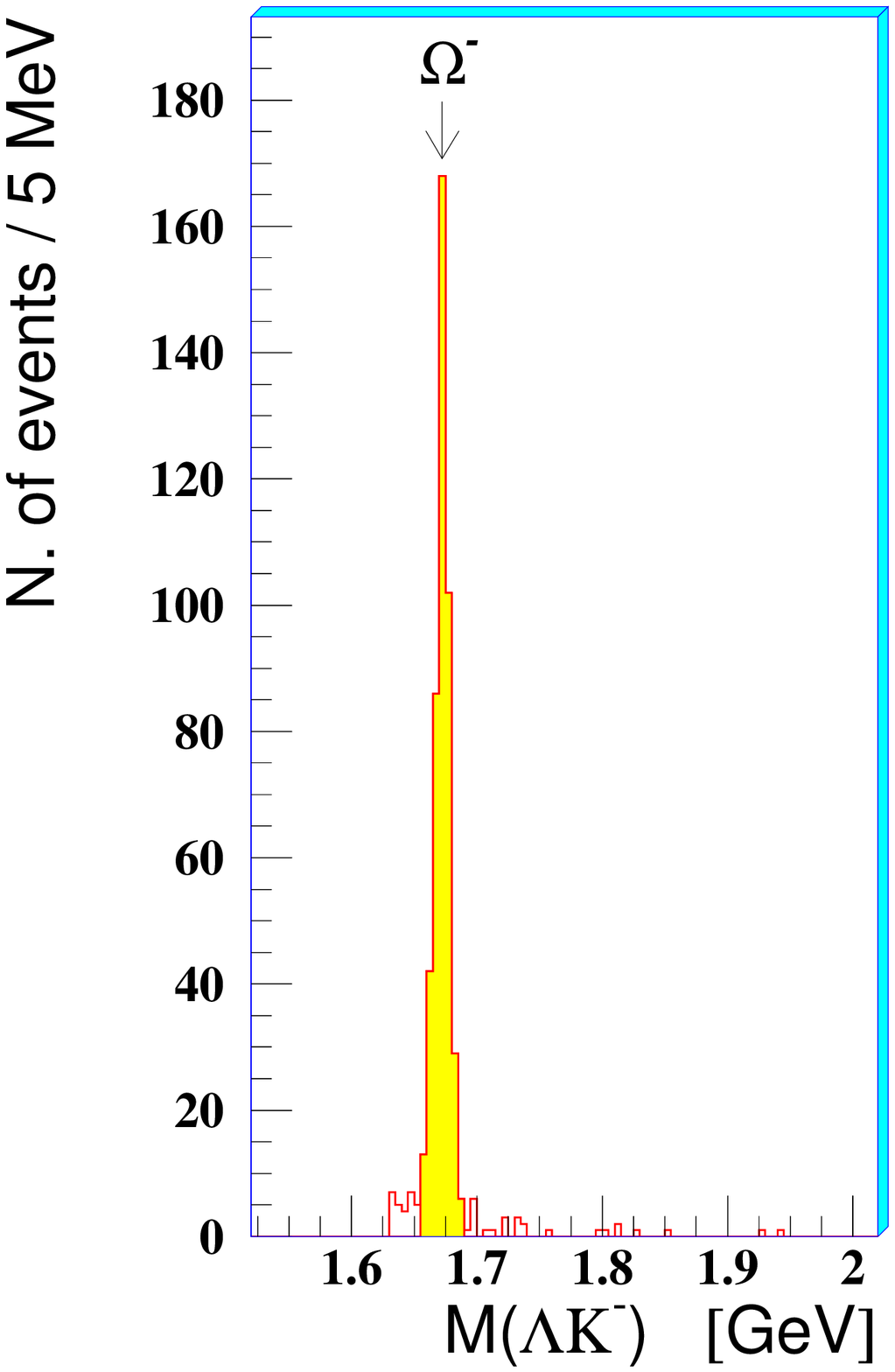}}
\resizebox{0.82\textwidth}{!}{%
\includegraphics{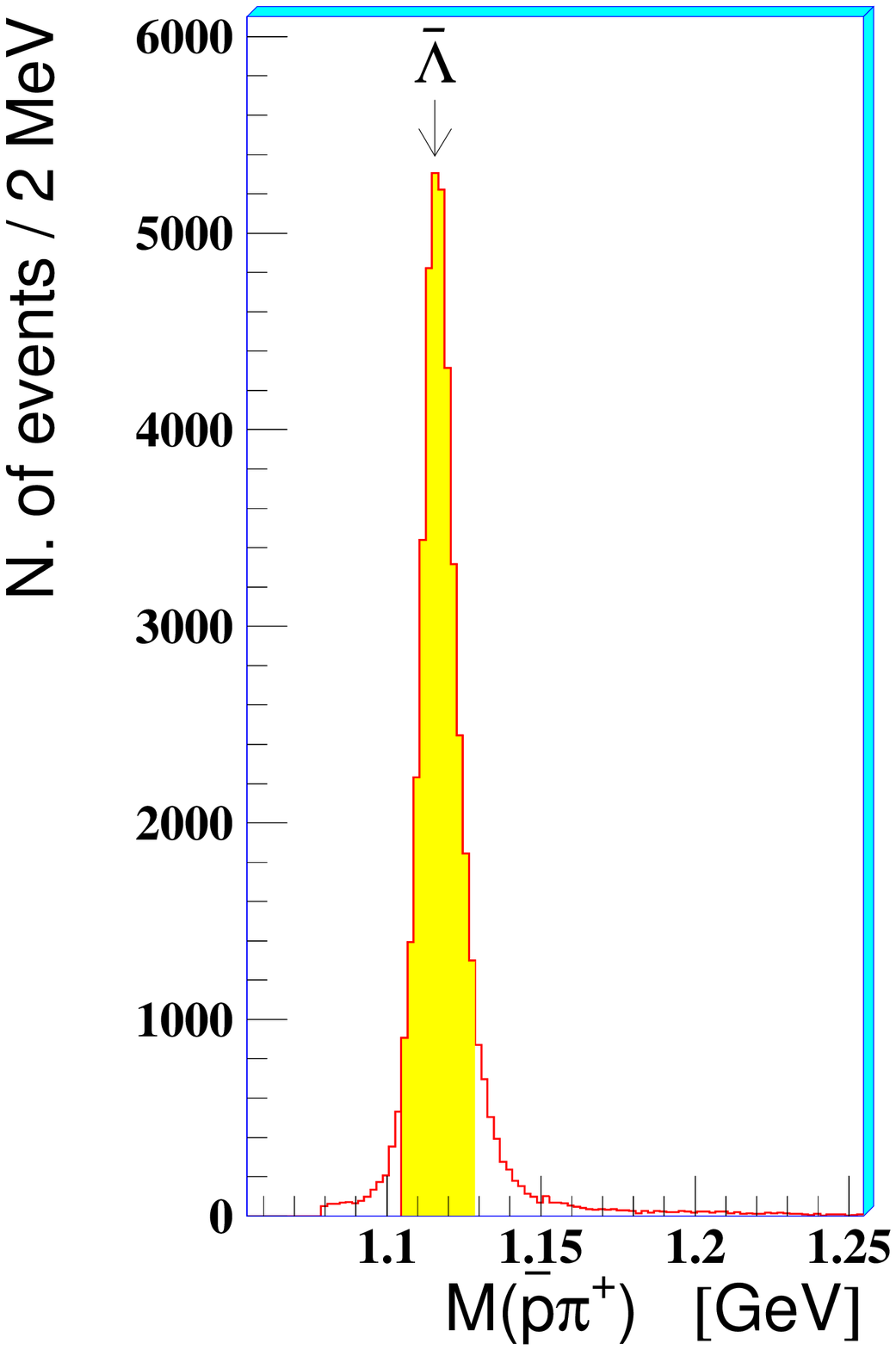}
\includegraphics{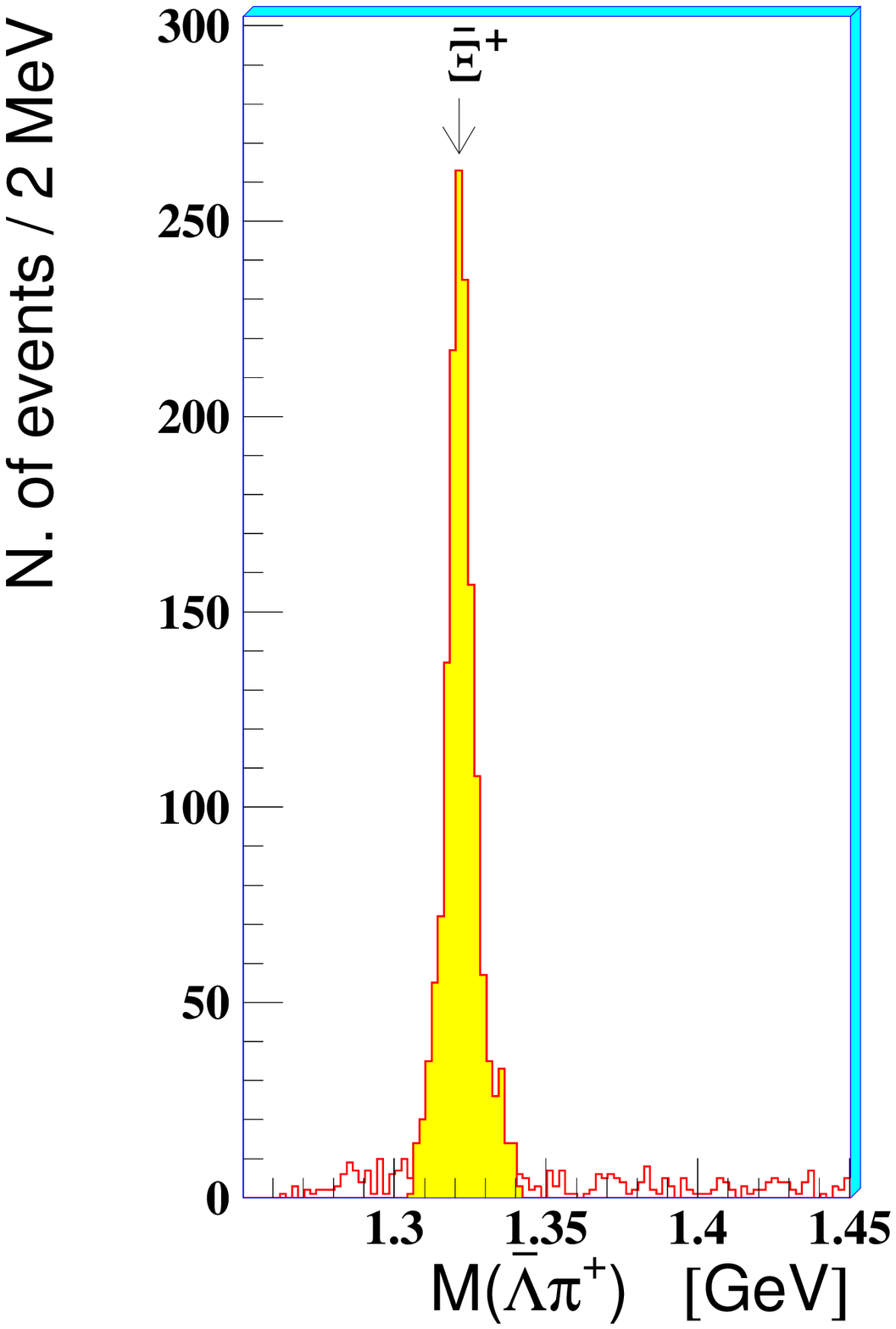}
\includegraphics{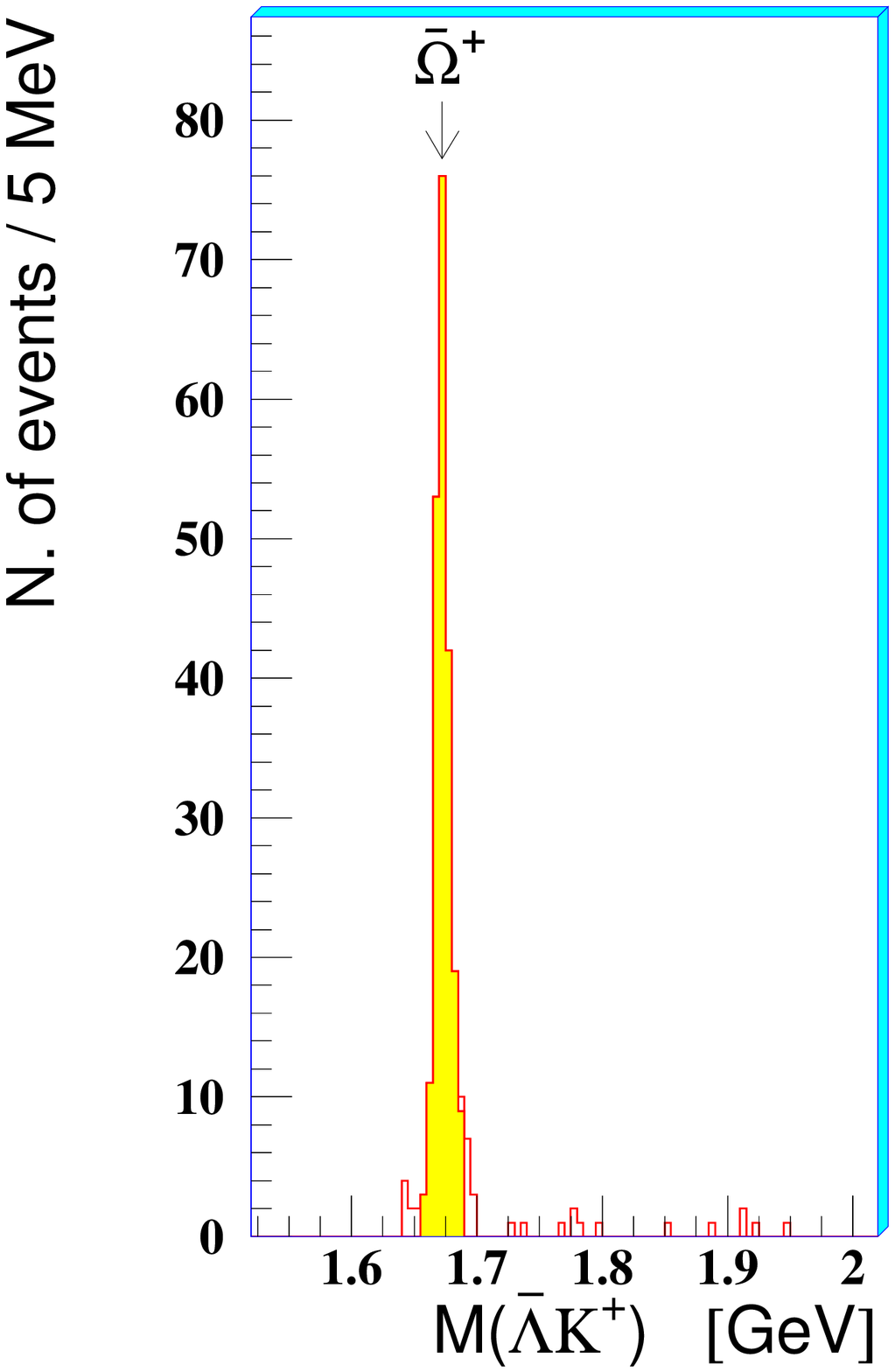}}
\caption{\rm Sample invariant mass spectra for \Pp\Pgp,
              \PgL\Pgp\ and \PgL\PK\ in Pb--Pb collisions.}
\label{3}
\centering
\resizebox{0.82\textwidth}{!}{%
\includegraphics{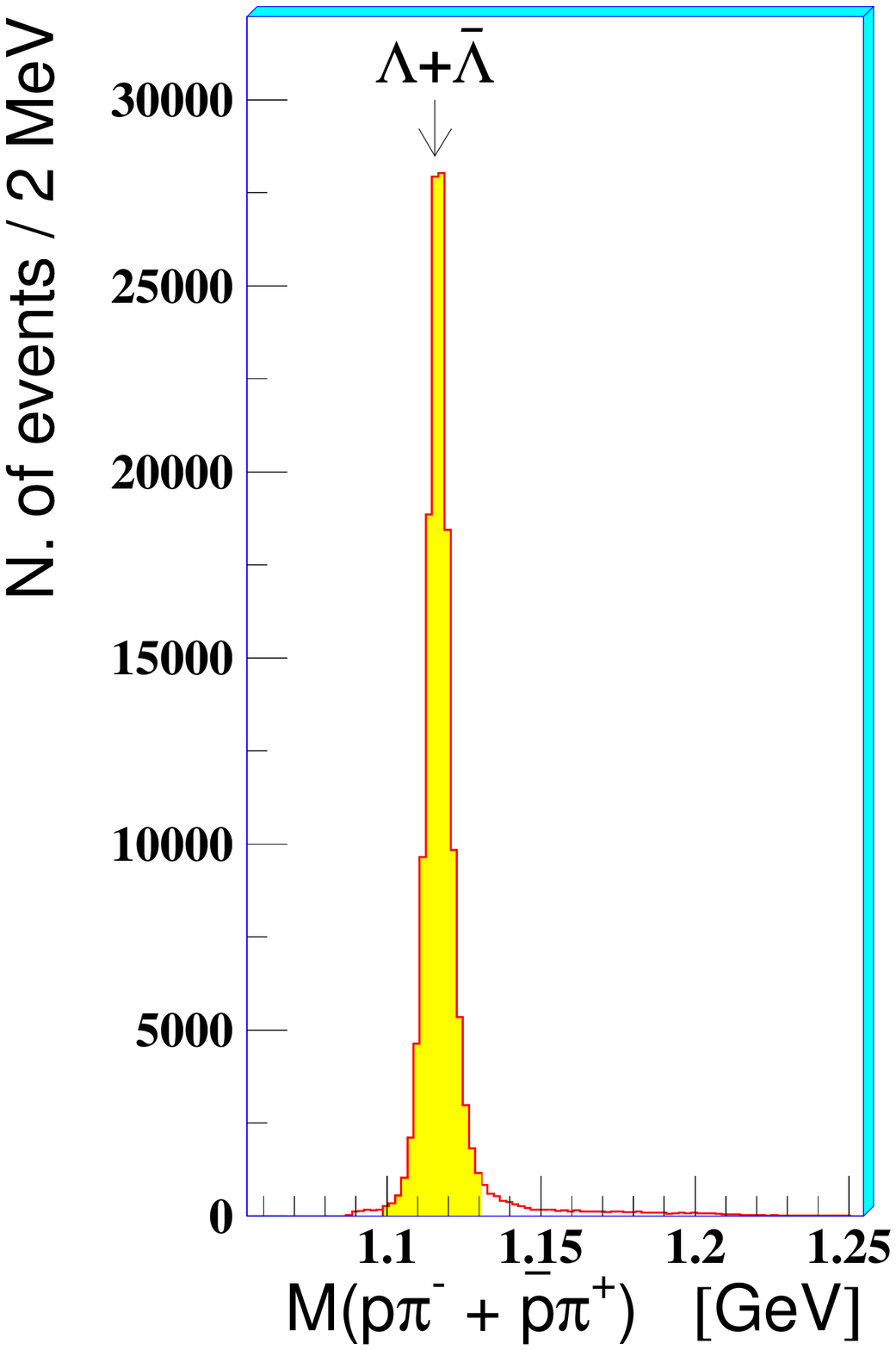}
\includegraphics{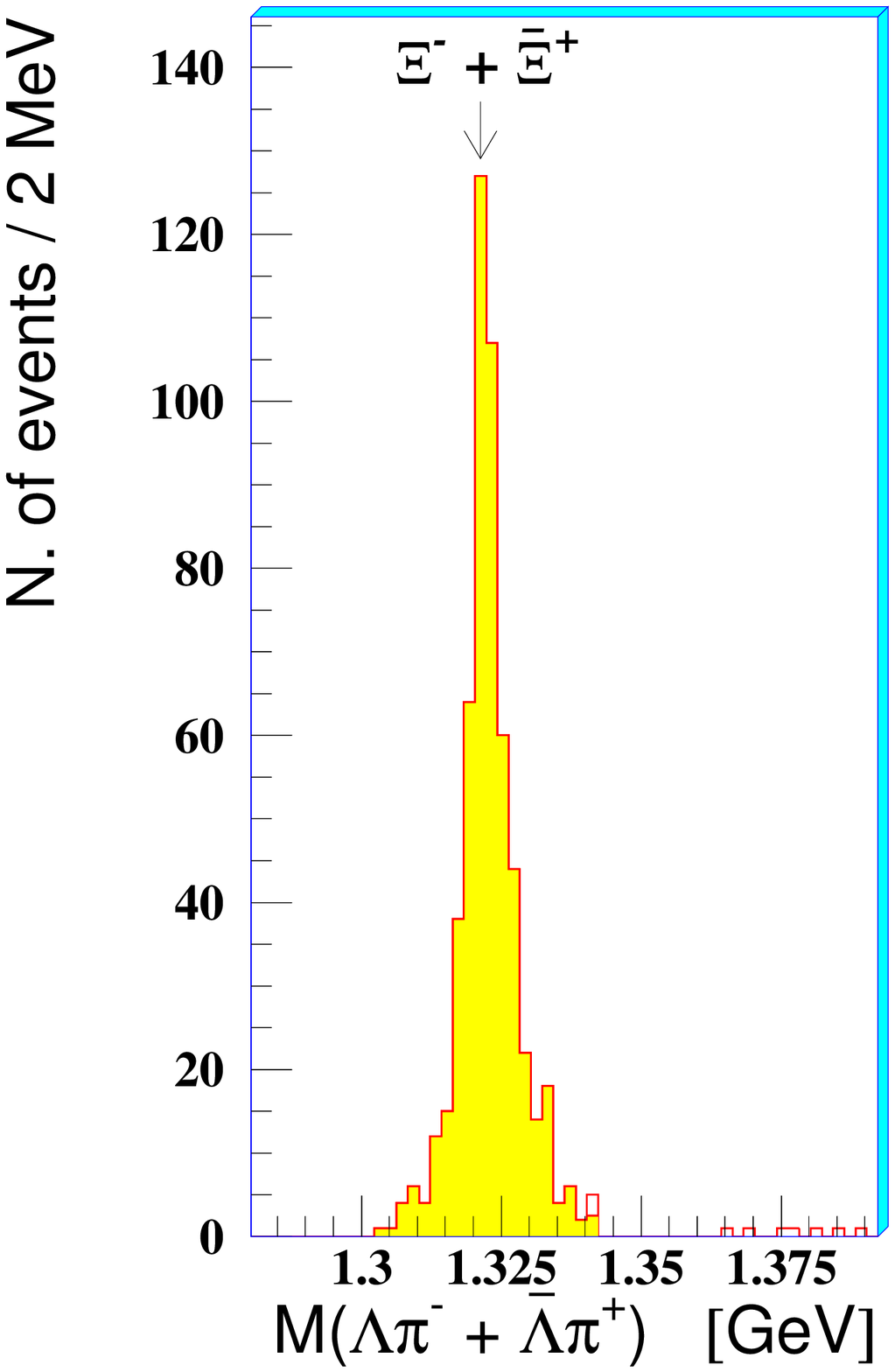}
\includegraphics{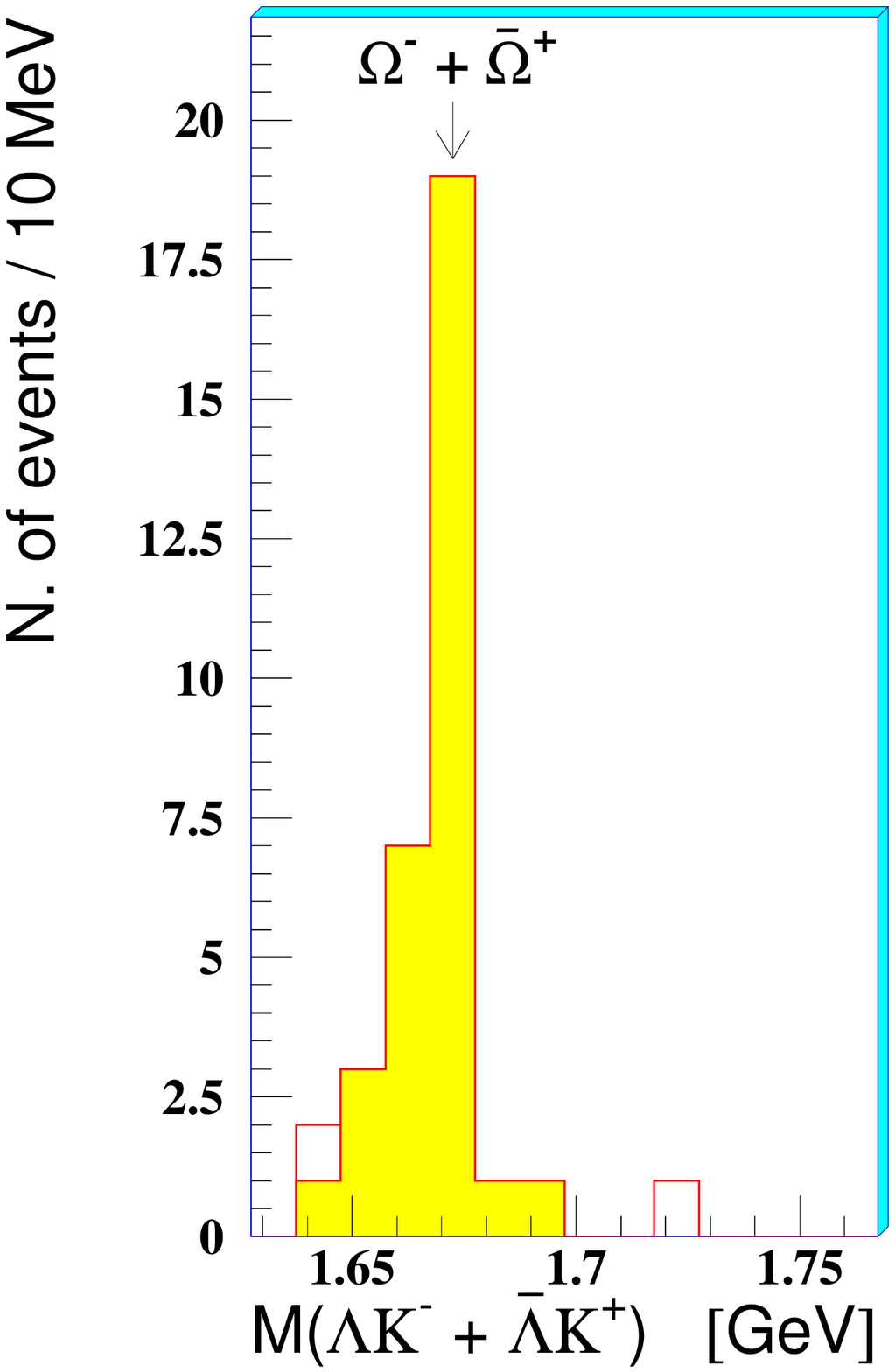}
}
\caption{\rm Sample invariant mass spectra for \Pp\Pgp,
              \PgL\Pgp\ and \PgL\PK\ in pBe collisions.
              
In all histograms the shaded area indicates the selected invariant mass intervals.}
\label{4}
\end{figure}

The quality of the NA57 data can be appreciated from figure~\ref{3}, where the invariant mass 
spectra of p$\pi$, $\Lambda\pi$, $\Lambda$K combinations obtained after applying all the 
selection criteria are displayed.  
\sublabon{figure}
\sublaboff{figure}
Hyperon peaks stand out over a small background; they are centred at 
the PDG values~\cite{pdg} with FWHM of about 10 MeV. The 
background level in figure~\ref{3} is very low; however a detailed study has been carried out,   
as explained in reference~\cite{giuseppe_mt}, in order to evaluate the residual  
combinatorial background. This has been found to be 0.3\% and 1.2\% for 
\PgL\ and \PagL\ respectively. For the cascade hyperons the background has been estimated 
to be less  than 4\% for $\Xi$\  
and less than 6\% for $\Omega$. It has been accounted for in the 
evaluation of the systematic errors. 

The corresponding plots for the pBe data are shown in figure~\ref{4}. The hyperon peaks are 
again well centred, with FWHM of about 6 MeV. 
Distributions of comparable quality were obtained for the pPb data.  
\subsection{Corrections for experimental biases}
\begin{figure}[hbt]
\centering
\resizebox{0.98\textwidth}{!}{%
\includegraphics{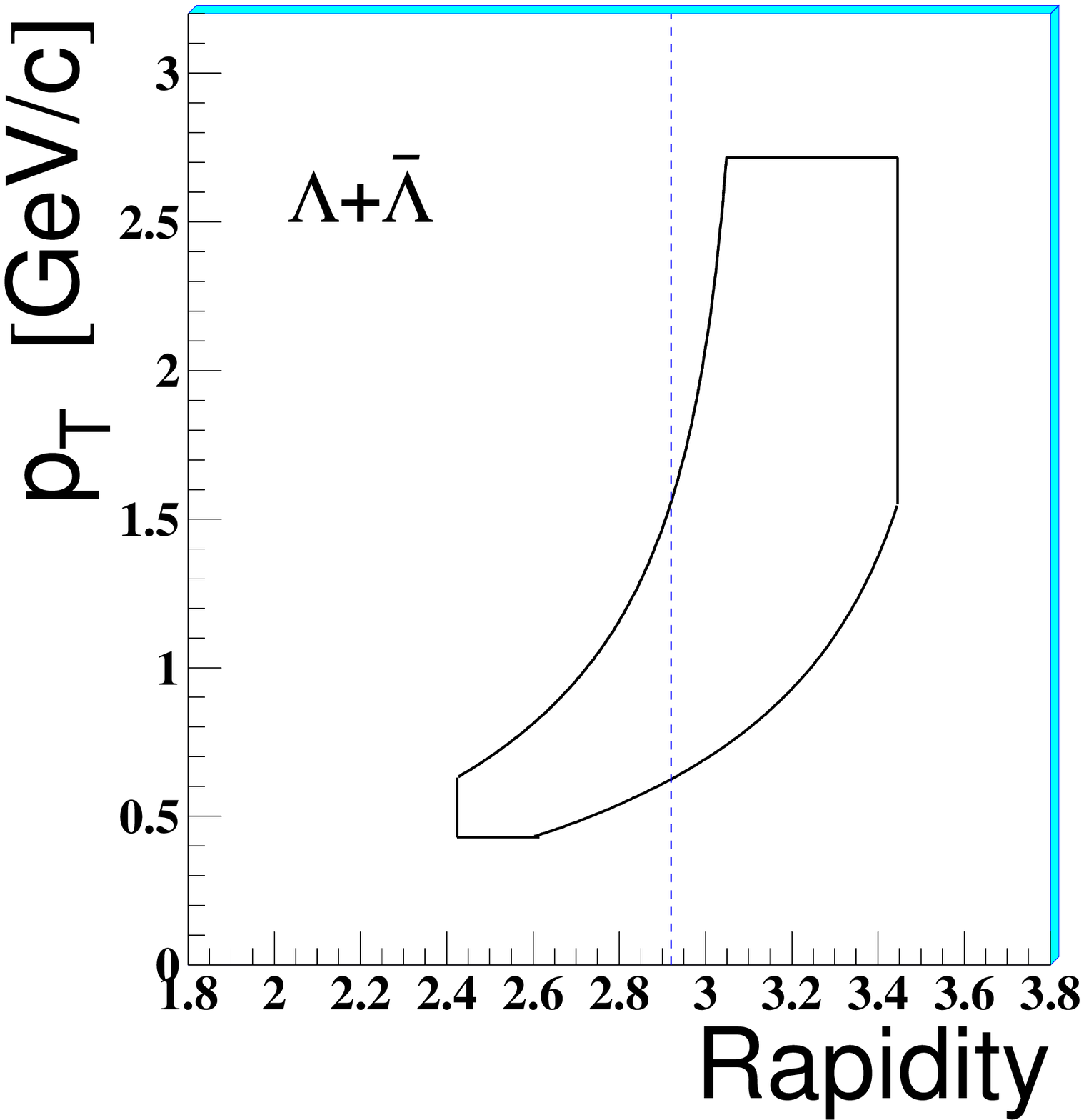}
\includegraphics{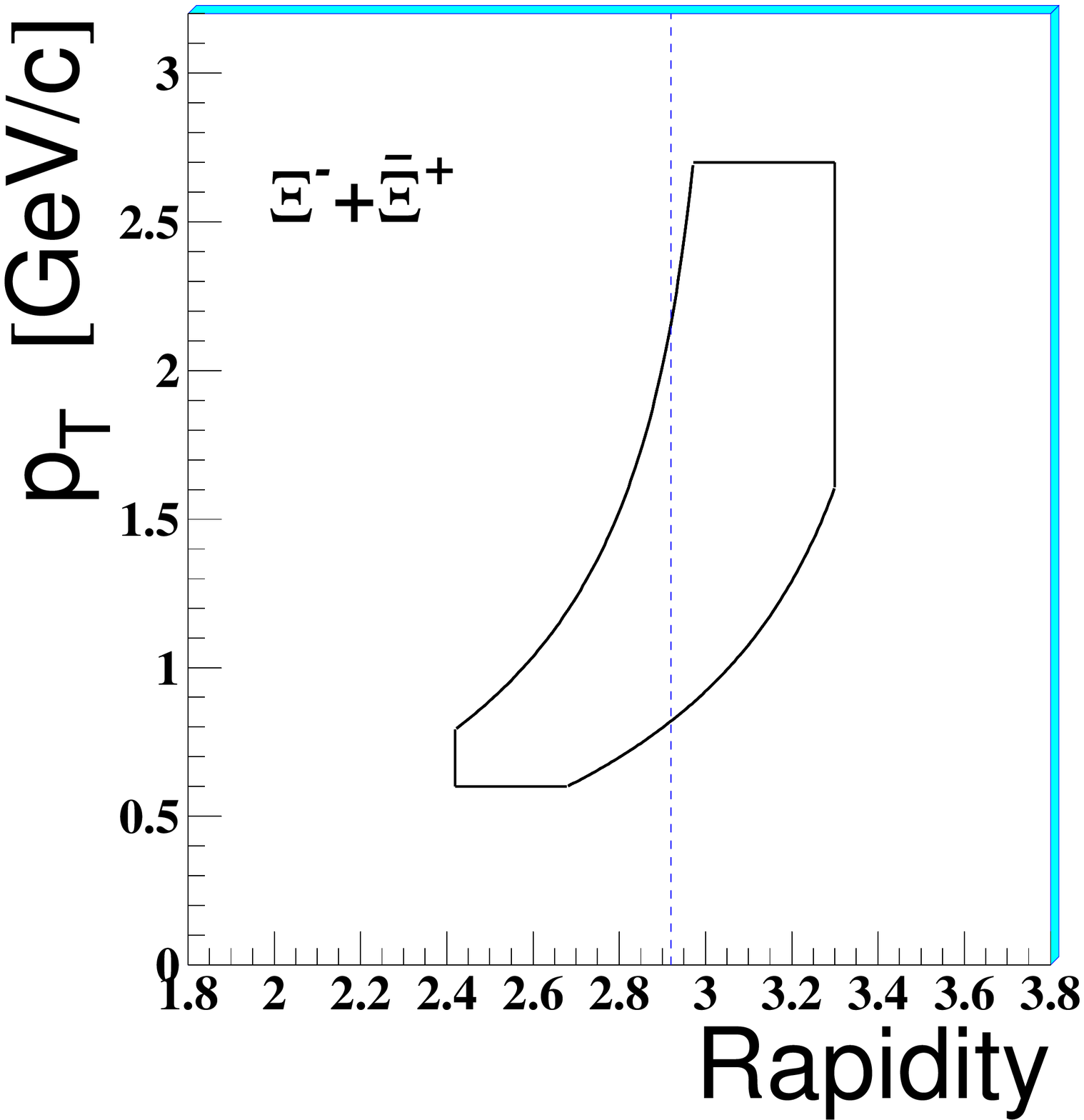}
\includegraphics{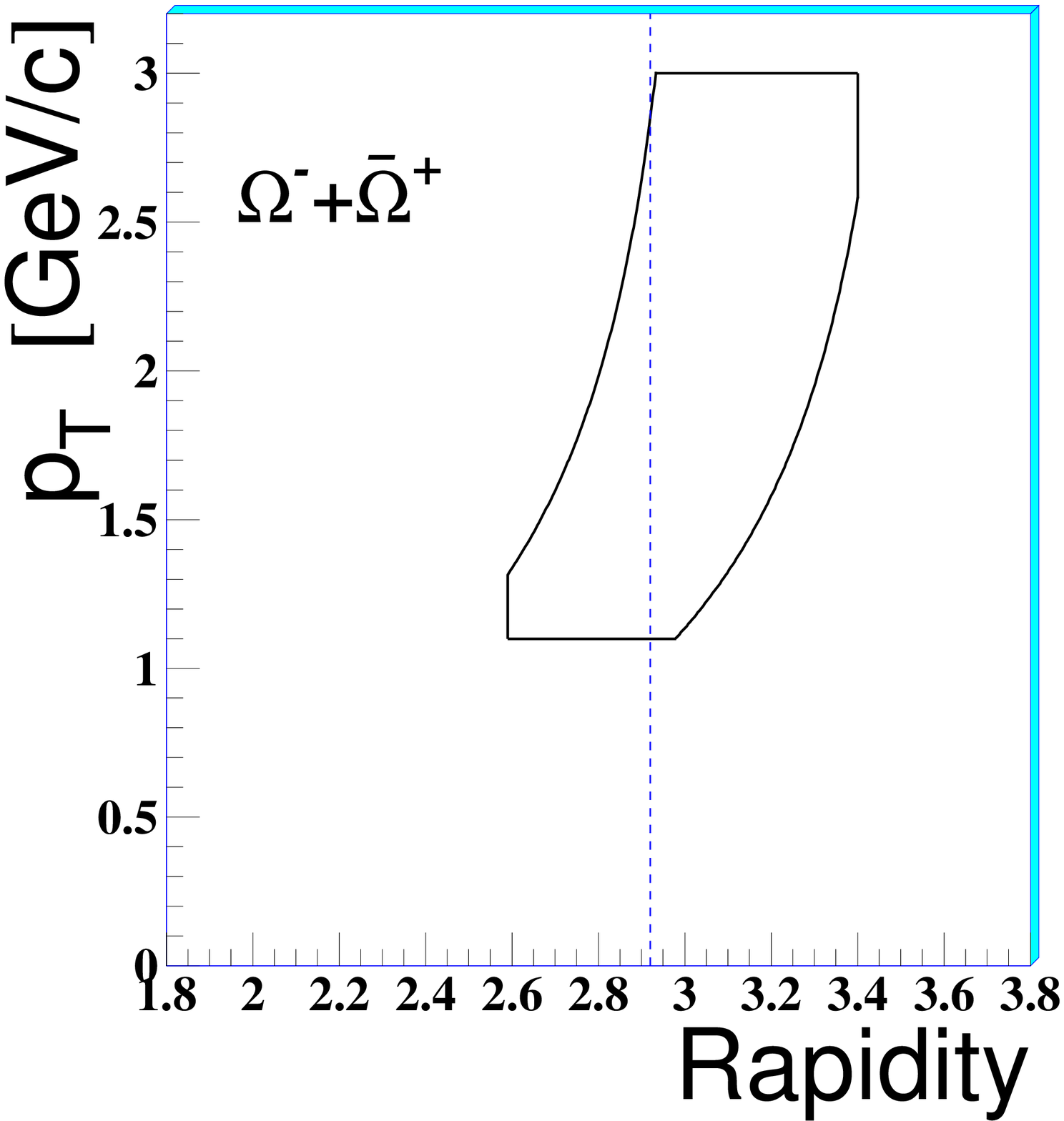}}
\caption{\rm The $y$--$p_{\tt T}$\
         acceptance windows 
         in Pb--Pb collisions.
         Dashed lines show the position of mid-rapidity ($y_{cm}=2.92$).}
\label{5}
\centering
\resizebox{0.98\textwidth}{!}{%
\includegraphics{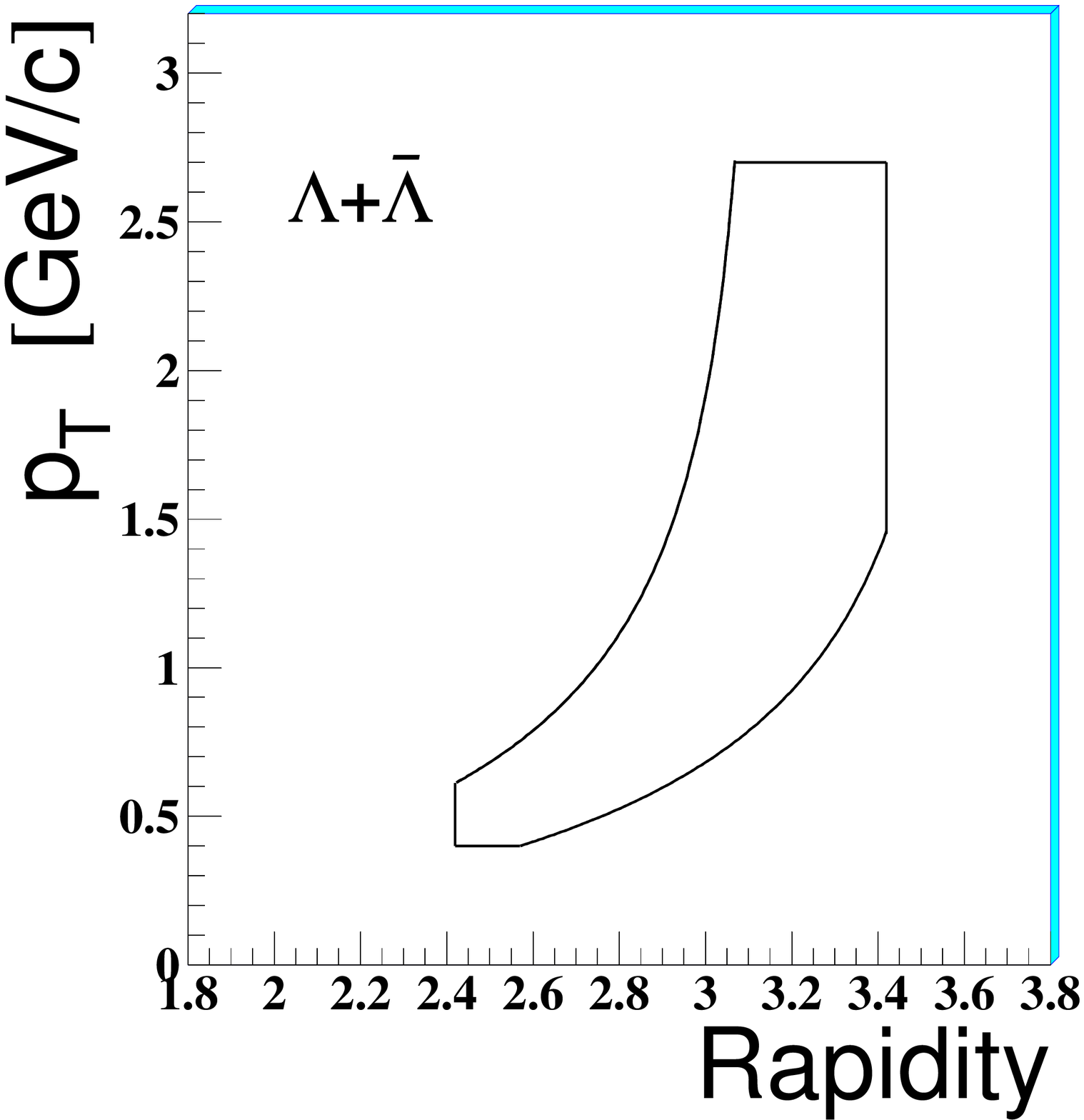}
\includegraphics{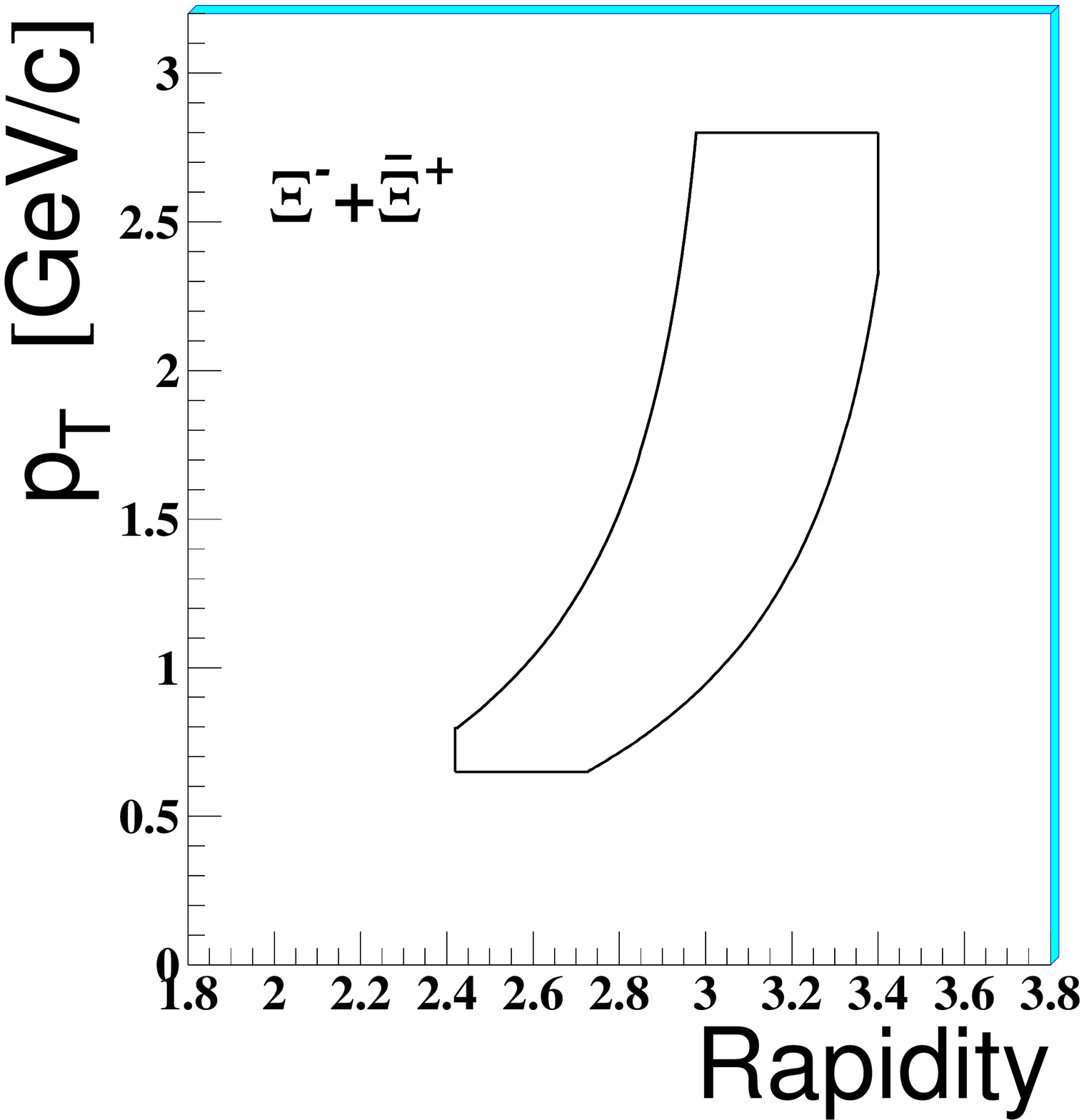}
\includegraphics{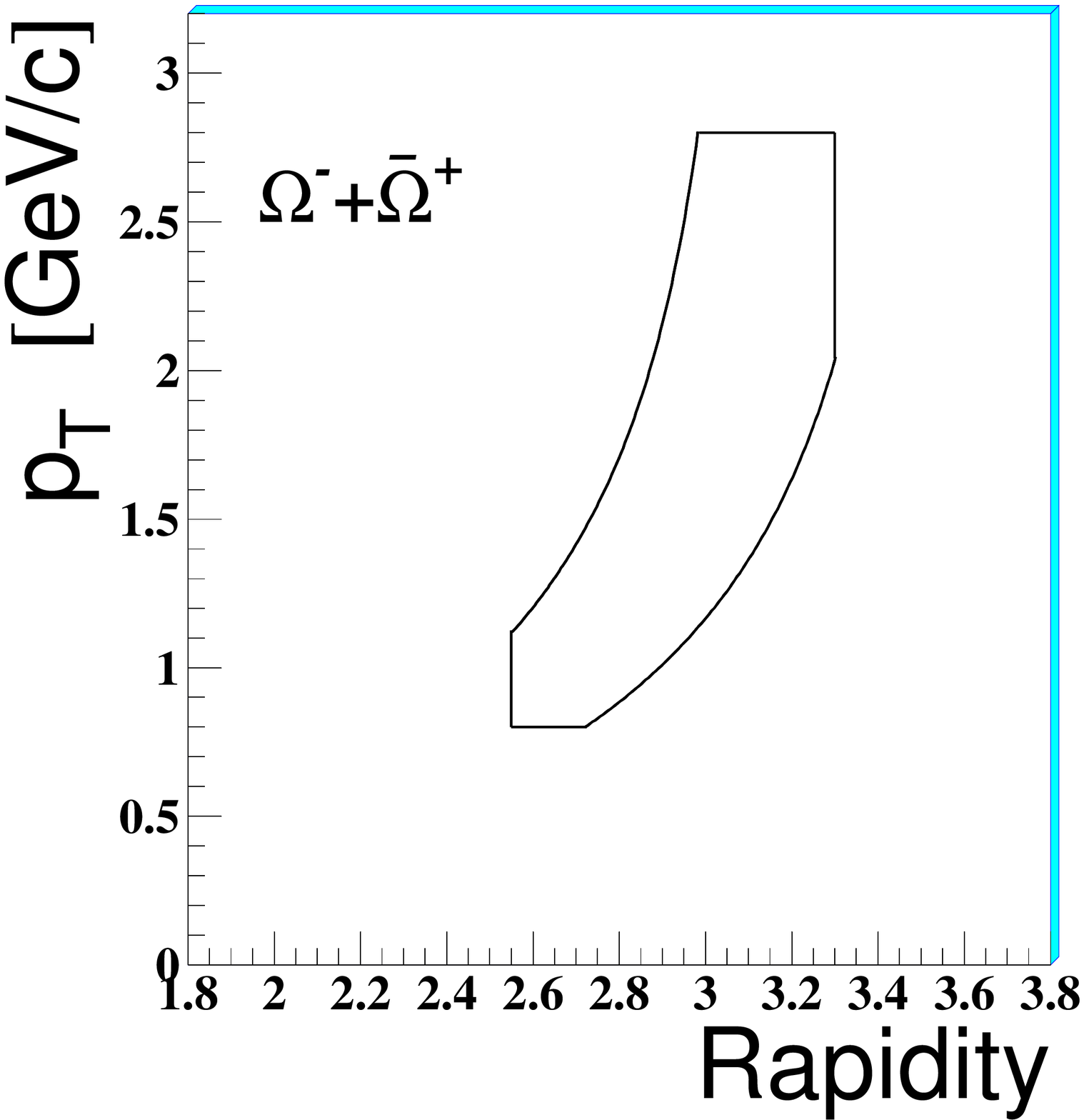}}
\caption{\rm The $y$--$p_{\tt T}$\  acceptance windows in pBe collisions.}
\label{6}
\end{figure}

The NA57 and WA97 experiments have been designed to accept particles 
produced at central rapidity ($y$) and medium $p_{\tt T}$. The exact limits of the acceptance region 
depend on the particle species; they have been refined off-line using a Monte Carlo simulation of the 
apparatus in order to define fiducial acceptance windows excluding the borders where the systematic 
errors are more difficult to evaluate. Sketches of the $y$--$p_{\tt T}$\ acceptance windows obtained for 
$\Lambda$, $\Xi$\ and $\Omega$\ in Pb--Pb and pBe interactions are shown in 
figure~\ref{5} and~\ref{6}, respectively. Those for pPb are similar to those for pBe interactions.  
\sublabon{figure}
\sublaboff{figure}


Data have been corrected for geometrical acceptance and for reconstruction and signal selection 
inefficiencies, using a Monte Carlo weighting procedure on a particle-by-particle basis.
For each reconstructed hyperon, the measured values of rapidity $y$ and transverse momentum $\pT$
have been used to generate a large sample of Monte Carlo particles with random azimuthal angle and  
the primary vertex position distributed according to the measured beam profile. The generated particles  
were then propagated through the NA57 set-up, simulated using the GEANT3 detector description and
simulation tool~\cite{Geant3}, and allowed to decay 
into the final states listed in equation~\ref{eq:decay} according to their lifetimes. 
The detector efficiencies have been determined individually for each read-out chip in the pixel planes 
and as an average for each microstrip plane, and were taken into account in the simulation.
                                                                                                              
For each Monte Carlo particle with all decay tracks passing the silicon telescope
the influence of background tracks and of the electronic noise was simulated by embedding
the hits generated by the Monte Carlo into a real event with hit multiplicity and reconstructed primary 
vertex position close to those of the original hyperon event. These mixed events were then  processed 
by the same program chain (pattern recognition, track reconstruction, hyperon reconstruction and 
selection) as the real data. The resulting weight assigned to the real event was calculated as the ratio 
of the number of generated Monte Carlo events to the number of Monte Carlo events successfully reconstructed and passing through all the analysis criteria.
An additional correction for the beam telescope efficiency has been applied for the pBe data.
                                                                                                            
The simulation program used for the calculation of the correction factors has been tested by 
comparing real and simulated distributions for all the main parameters used in the selection of the signals. 
A good agreement of real and simulated spectra has been found \cite{giuseppe_mt,kristin_02}. 

We have calculated weights for all the reconstructed multi-strange particle candidates. 
However, the weighting procedure is very CPU intensive and therefore for the 
much more abundant $\Lambda$ and $\overline\Lambda$ 
samples we only corrected a fraction of the total data 
in order to reach a statistical precision better than the limits imposed by the systematics.

Given the geometry of the two experiments the feed-down from weak decays
is of minor importance. It 
has been estimated to be less than 5\% for  
\PgL\ from \PgXm\ and less than 10\% for \PagL\ from \PagXp; for both \PgXm\  
and \PagXp the feed-down from $\Omega$\ decays is less than 2\%. 

\subsection{Centrality determination}
\begin{figure}[h!]
\begin{center}
 \resizebox{0.52\textwidth}{!}{%
 \includegraphics{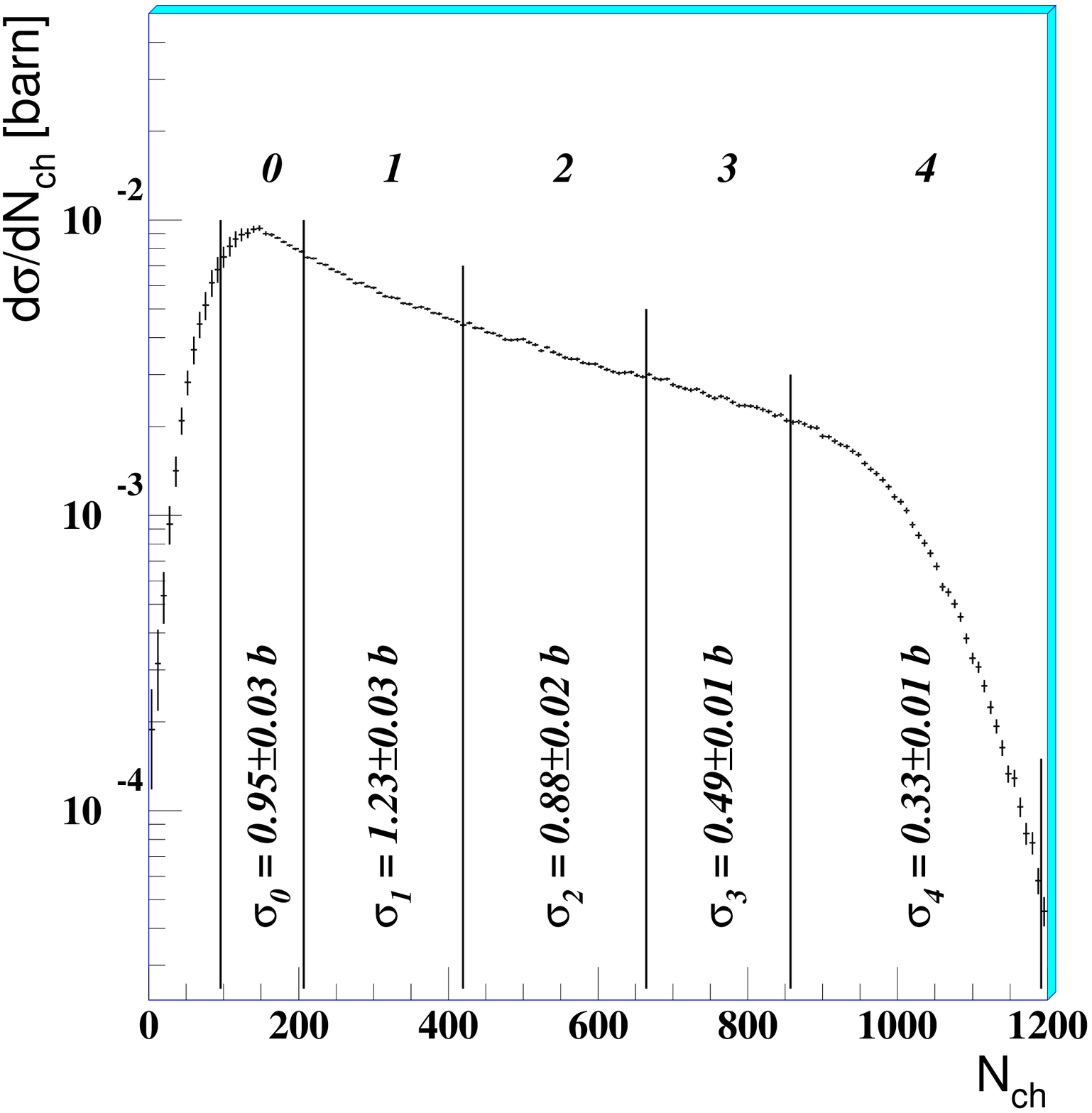}}
\end{center}
\caption {\label{mult_bins} Charged particle multiplicity distribution
          and centrality classes for 158 $A$ GeV/$c$ Pb--Pb collisions.}
\end{figure}

As a measure of the collision centrality we use the number of wounded nucleons
$N_{\rm{wound}}$~\cite{glauber,bialas} extracted from the charged 
particle multiplicity measurement.
The details of the multiplicity detector and reconstruction of the charged 
multiplicity distribution are described in~\cite{nicola_c,tiziano}.
Figure~\ref{mult_bins} shows the charged particle multiplicity distribution for Pb--Pb collisions at 158 
$A$ GeV/$c$ as measured by the NA57 experiment.
The drop at low multiplicities is due to the threshold of the centrality trigger. The data have been binned 
in five classes as indicated in figure~\ref{mult_bins}. The events with multiplicity below class 0 have 
been excluded from the subsequent analysis; the final sample corresponds to the most central 
53\% of the Pb--Pb inelastic cross-section. The distribution of
$N_\mathrm{wound}$ for the five multiplicity classes was computed from the measured trigger 
cross sections and a modified Wounded Nucleon Model fit to the multiplicity distribution using the 
Glauber model~\cite{glauber, nicola_c}. The fractions of inelastic cross section for the selected 
classes, as well as the corresponding average numbers of wounded nucleons 
$\langle N_{\mathrm{wound}}\rangle$ are shown in table \ref{tab:centrality}. 

\begin{table}[h]
\caption{Centrality ranges and values of
$\langle N_{\mathrm{wound}}\rangle$ for the selected classes. 
Statistical (first) and systematic (second) errors are quoted.
\label{tab:centrality}}
\begin{center}
\begin{tabular}{cccccc}
\hline
 Class &   $0$   &   $1$   &   $2$   &  $3$   &   $4$ \\ \hline
 $\sigma/\sigma_{inel}$\ \; (\%)   & 40 to 53 & 23 to 40 & 11 to 23 & 4.5 to 11 & 0 to 4.5 \\
 $\langle N_{\mathrm{wound}}\rangle$ & $62 \pm 4 \pm 6$ & $121 \pm 4 \pm 7$ & 
  $209 \pm 3 \pm 6$ & $290 \pm 2 \pm 4$ & $349 \pm 1 \pm 1.4$ \\ 
 \hline
\end{tabular}
\end{center}
\end{table}

The number of wounded nucleons in pBe collisions ($\langle N_{\mathrm{wound}}\rangle = 2.5$) 
and in pPb collisions ($\langle N_{\mathrm{wound}}\rangle = 4.75$) have been determined 
from a Glauber model calculation  
as an average over all inelastic collisions. 

\section{Particle yields and strangeness enhancements}
The double differential cross sections for each particle under study were fitted using
the expression
  \begin{equation}
   \frac{{\rm d}^2 N}{\dder\mT\,\dder y}
    = f(y)\,\mT\,\exp\left(-\frac{\mT}{T}\right) ,
  \label{ymtdist}
  \end{equation}
where $\mT=\sqrt{m^2+p_\mathrm{T}^2}$  is the transverse mass and $y$ is the rapidity.
The inverse slope $T$\ is interpreted as an apparent temperature 
due to the thermal motion 
coupled with a collective transverse flow of the fireball components
~\cite{giuseppe_mt,NA44Bearden}. 
The fit was performed using the method of maximum likelihood, keeping $T$ as a free parameter. 
The complete results concerning the inverse slopes  
from the NA57 and WA97 experiments are given in references~\cite{giuseppe_mt} and~\cite{fini_mt}, 
respectively.

For the present analysis 
we assumed a flat rapidity distribution ($f(y) = {\rm const}$) in our acceptance region  
for all particles except \PagL\ in Pb--Pb; 
the rapidity distribution of the \PagL\ 
is significantly non-flat within our rapidity window and 
it has been parametrised as a Gaussian~\cite{rapidity}.  

Using the parametrisation of the differential cross section given by equation
\ref{ymtdist} with the inverse slope values obtained from fits, we have determined  
the yield Y of each particle under study, extrapolated to a common phase space region
covering the full $\mT$ range and one unit of rapidity centred at mid-rapidity $y_{\rm cm}$:
\begin{equation}
Y =\int^{\infty}_m \dder\mT \int^{y_{cm}+0.5}_{y_{cm}-0.5}
\frac{\dder^2 N}{\dder\mT\,\dder y} ~\dder y \; .
\label{ex_yield}
\end{equation}

Extrapolated yields were calculated for each centrality class (0--4) of Pb--Pb collisions.   
The resulting values are presented in table \ref{vm_qm02}.
We also report the yields for the pBe and pPb interactions  
measured in the WA97 experiment.
The assumption of a flat rapidity distribution, in the case of asymmetric collisions (p--nucleus)  
induces a systematic error on the particle yields which is at most of  order  
5-6~\%~\cite{wa97_pl2}.   

\begin{table}[h]
\caption{Strange particle yields in Pb--Pb for the five centrality classes (top) and
         for pBe and pPb collisions (bottom). The first error is statistical, the second one 
is systematic. \vspace{0.3cm}
\label{vm_qm02}}
\footnotesize{
\hspace{-0.6cm}
\begin{tabular}{|l|ccccc|}
\cline{2-6} \multicolumn{1}{c|}{ }
                     &   0      &    1      &    2      &    3    &    4   \\ \hline
\PgL
      & $2.30\pm0.22\pm0.23$ & $5.19\pm0.29\pm0.51$ & $9.5\pm0.5\pm0.9$ & $15.0\pm0.8\pm1.5$ & 
        $18.5\pm1.1\pm1.8$\\ \hline
\PagL &$0.41\pm0.03\pm0.04$ &$0.80\pm0.04\pm0.08$ &$1.58\pm0.07\pm0.16$&$1.81\pm0.10\pm0.18$&
       $2.44\pm0.14\pm2.4$\\ \hline
\PgXm  
 & $0.181\pm0.013\pm0.018$ & $0.52\pm0.02\pm0.05$ & $1.07\pm0.04\pm0.11$ & $1.80\pm0.07\pm0.18$ &
   $2.08\pm0.09\pm0.21$  \\ \hline
\PagXp &
       $0.045\pm0.007\pm0.004$ &$0.14\pm0.01\pm0.01$&$0.29\pm0.02\pm0.03$&$0.37\pm0.03\pm0.04$&
       $0.51\pm0.04\pm0.05$\\ \hline
%
\PgOm\ &  
$0.024\pm0.016\pm0.004$ & $0.064\pm0.013\pm0.010$ & $0.17\pm0.03\pm0.03$ & $0.22\pm0.04\pm0.03$ &
$0.31\pm0.07\pm0.05$ \\ \hline
\PagOp &  
$0.013\pm0.008\pm0.002$ & $0.031\pm0.010\pm0.004$ & $0.064\pm0.015\pm0.010$ & $0.11\pm0.03\pm0.02$ &
$0.16\pm0.04\pm0.02$ \\ 
\hline
\end{tabular}
%
%
\begin{center}
\vspace{0.4cm}
\begin{tabular}{|l|cc|}
\cline{2-3} \multicolumn{1}{c|}{ }
                     &   pBe      &    pPb   \\ \hline
%
%
\PgL &  $0.0334\pm0.0005\pm0.003$   &    $0.060\pm0.002\pm0.006$  \\ \hline
\PagL &   $0.0111\pm0.0002\pm0.001$   &    $0.015\pm0.001\pm0.002$\   \\ \hline
\PgXm &  $0.0015\pm0.0001\pm0.0002$   &    $0.0030\pm0.0002\pm0.0003$  \\ \hline
\PagXp &  $0.0007\pm0.0001\pm0.0001$ &    $0.0012\pm0.0001\pm0.0001$  \\ \hline
%
\PgOm &  $0.00012\pm0.00006\pm0.00002$ & $0.00022\pm0.00008\pm0.00003$  \\ \hline 
\PagOp&  $0.00004\pm0.00002\pm0.00001$ & $0.00005\pm0.00003\pm0.00001$  \\ \hline 
\end{tabular}
\end{center}
}
\end{table}

We have performed a number of stability tests, changing the choice of the acceptance region or the 
set of the analysis cuts, and analysing separately data samples taken with opposite magnetic field 
polarities and data samples recorded in different data taking periods~\cite{kristin_02,giuseppe_tesi}. 
This study  
demonstrated good stability 
for the procedure for extracting yields.  
The overall systematic errors, both for particle yields and transverse mass distributions, are estimated 
to be about 10\% for $\Lambda$ and $\Xi$ and about 15\% for $\Omega$. 

Centrality classes 1 to 4 correspond to those of the WA97 experiment. 
In this common centrality range the yield values from the NA57 and WA97 experiments agree
within 20-30\%. This difference has been investigated \cite{kristin_02} and 
has been understood as due mainly to a bias introduced by the treatment of the
uncertainty in the position of the primary vertex in WA97. This comes about because of fluctuations 
in the beam-position.
This bias has not affected the pBe and pPb data due to the greater stability of the proton beam; in addition,
in the case of pBe the beam position was accurately measured with the beam telescope.

Using the extrapolated yields we have determined the strangeness enhancement E,
which we define as the yield per participant in Pb--Pb (or pPb) collisions normalised to 
the yield per participant in pBe collisions:
\begin{equation}
 E = \left(\frac{Y}{\langle {N_{\mathrm{wound}}} \rangle } \right)
     /
     \left(\frac{Y}{\langle {N_{\mathrm{wound}}} \rangle } \right)_{\mathrm{pBe}} 
\label{enhance}
\end{equation}

\begin{figure}[t!]
\centering
 \resizebox{0.60\textwidth}{!}{%
 \includegraphics{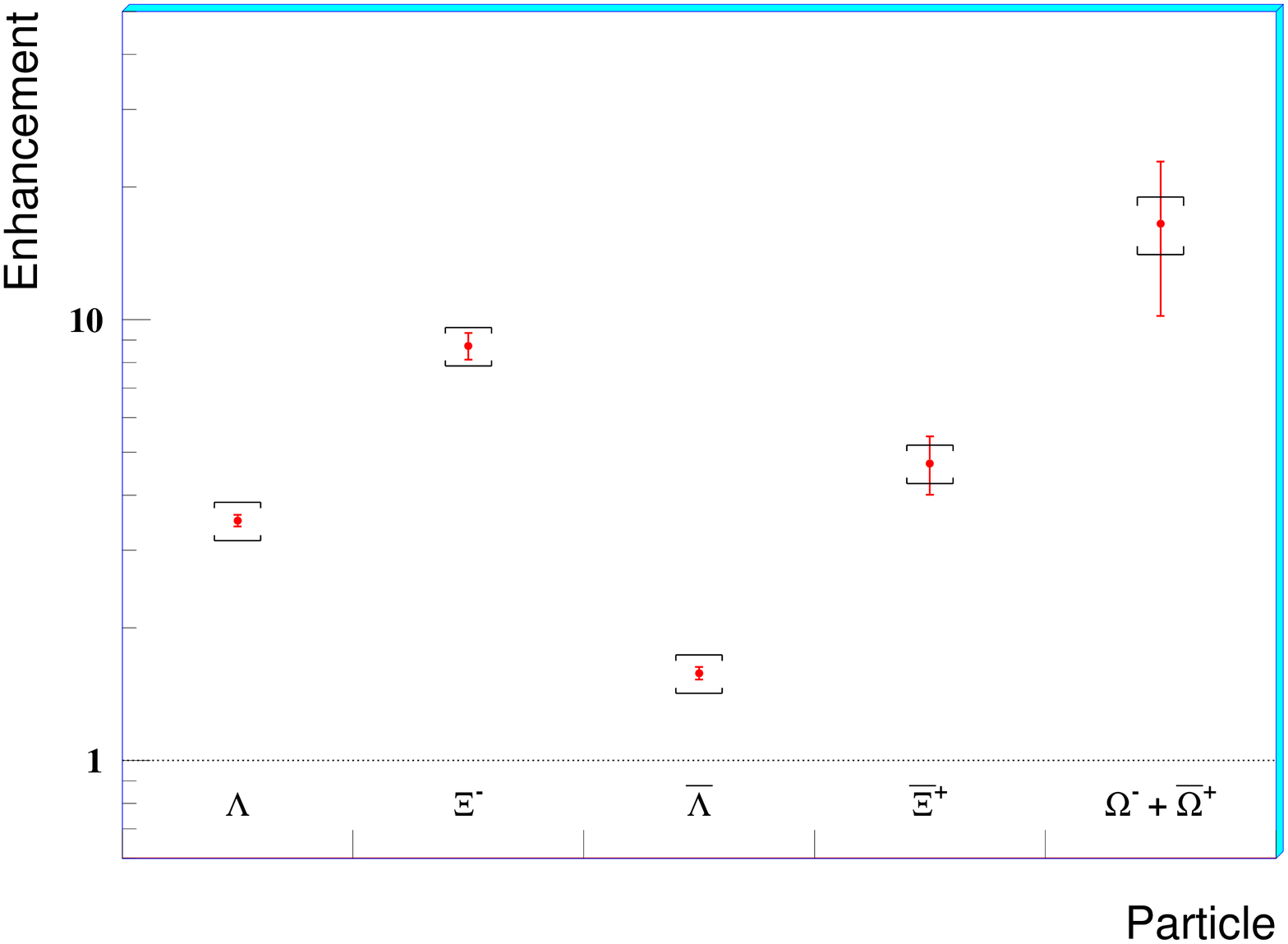}}
 \caption {\label{enh_globalb} Hyperon enhancement versus strangeness content for 53\% most
central Pb--Pb events. 
Errors on the pBe yields have been propagated. 
The symbol $^\sqcap_\sqcup$ shows the systematic error.  
Due to the limited statistics in the  pBe sample the \PgOm and \PagOp signals have been combined.}   
\end{figure}

The hyperon enhancements in the full centrality region (53\% most central Pb--Pb events) 
are shown in figure~\ref{enh_globalb} 
as functions of the strangeness content of the particle. A clear hierarchy of the enhancements is 
visible: they increase with the strangeness content of the particle, for both hyperons and 
antihyperons, reaching a factor of about 20 for the triply-strange $\Omega$.

\begin{figure}[h!]
\centering
  \resizebox{0.90\textwidth}{!}{%
 \includegraphics{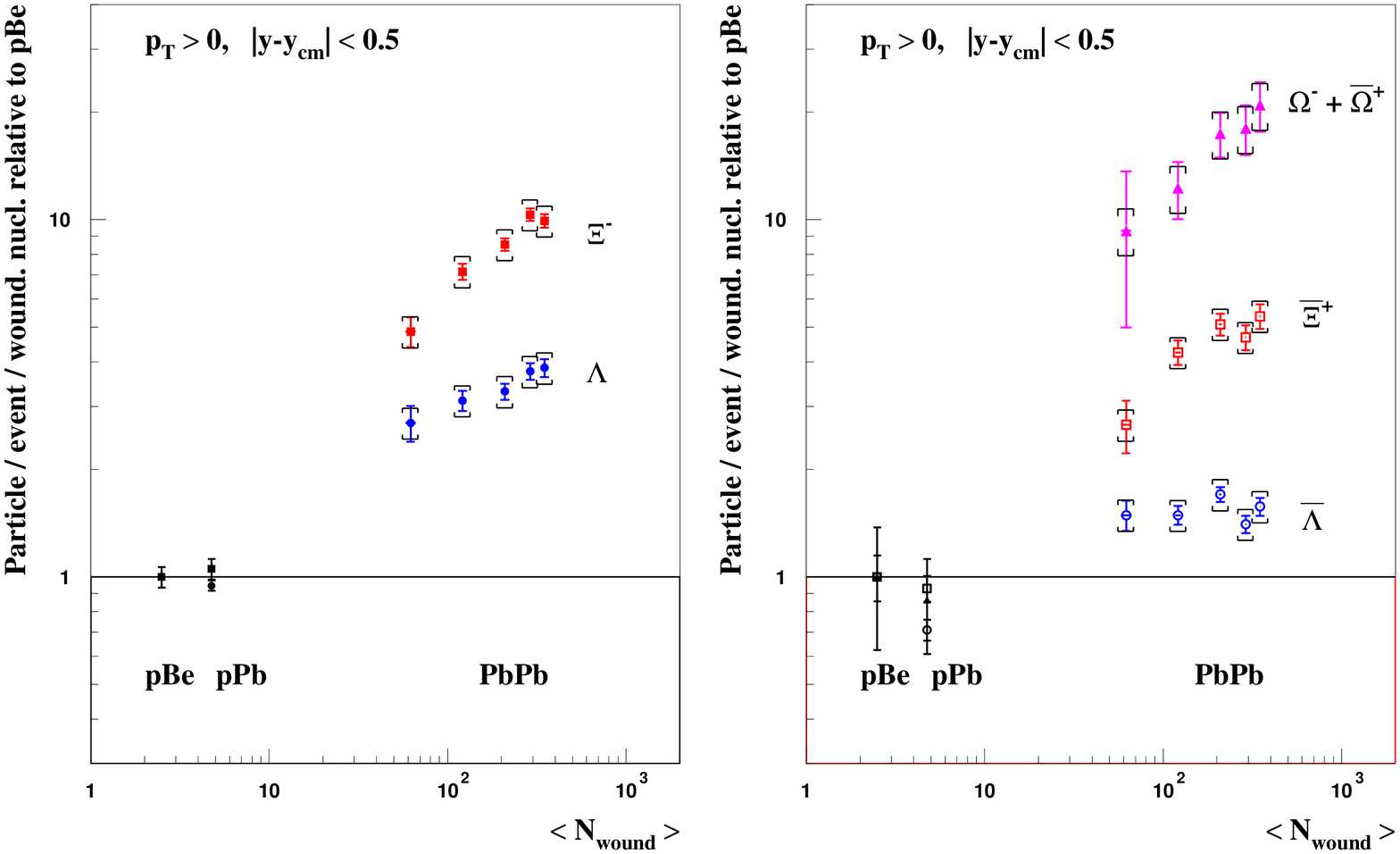}}
 \caption {\label{new_enh} Hyperon enhancements {\it E} as a function of the number
           of wounded nucleons. The symbol $^\sqcap_\sqcup$ shows the systematic
           error.  
}
\end{figure}

The behaviour of the  enhancements with the centrality of the collision is displayed in figure \ref{new_enh}, where the NA57 data for the five centrality classes defined above are plotted 
as a function of the number of wounded nucleons, together with the pBe and pPb results from the WA97 experiment.
The enhancements are shown separately for particles containing at least one valence
quark in common with the nucleon (left) and for the other particles (right), kept separate since 
in principle they may have different production features.
   
It is worth noting that the normalized yields for the pBe and pPb data are 
compatible with each other within the error limits, as expected from $N_{\rm wound}$ 
scaling.\footnote{For a discussion of the $N_\mathrm{wound}$ scaling in pA see reference~\cite{antinori_qm04}.} 
The Pb--Pb data exhibit a significant centrality dependence of the enhancements for all particles except $\overline\Lambda$. However, for the two most central classes
3 and 4 ($\simeq 10\%$ of most central collisions) a saturation of the enhancements
cannot be ruled out, in particular for \PgXm and \PagXp. 

\section{Discussion}
The results presented above refine and extend the study of strangeness
enhancements initiated by the WA97 experiment~\cite{wa97_pl2,wa97_qm99}. 
In particular  the increase of the magnitude of the enhancement with the strangeness 
content of the particles is confirmed, as expected in a QGP scenario~\cite{rafelski}.  
The highest enhancement is measured for the triply-strange $\Omega$ hyperon 
and amounts to about a factor 20 in the most central class. 

\begin{figure}[h!]
\centering
  \resizebox{0.90\textwidth}{!}{%
 \includegraphics{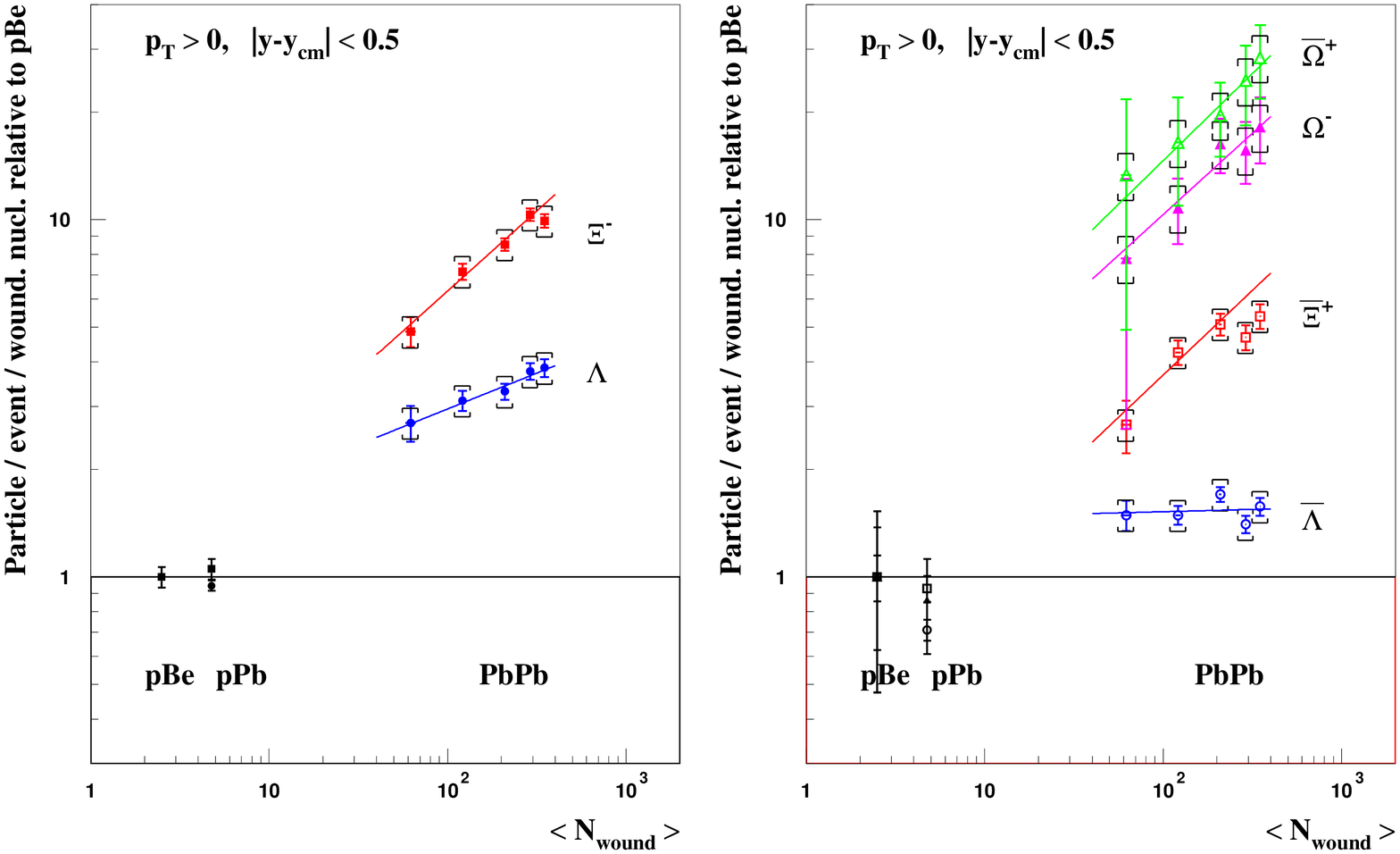}}
 \caption {\label{enh_fit_few} Power law fit to the hyperon enhancements. 
 Statistical errors only are used in the fits. 
 For the \PagXp distribution, the last two points have been excluded from the fit.  
 For the \PgXm\ distribution, the last point only has been dropped; however, 
 excluding from the fit one or two points makes no appreciable differences in the parameters.}
 \end{figure}

In order to study the behaviour of the enhancements with the centrality of the collision, we have 
fitted the Pb--Pb data of figure \ref{new_enh} with a power law 
$E = A \langle N_{\rm wound}\rangle$$^{b}$
Good fits were obtained for all particles except $\PgXm$ and $\PagXp$, for which a possible 
saturation in the most central classes 
cannot be ruled out, as mentioned above.  
Excluding from the fit the two most central values  
gives good $\chi^{2}$  
also for the $\PgXm$ and $\PagXp$ distributions. 
The results of the fits are shown in figure \ref{enh_fit_few} superimposed on 
the enhancement distributions; the values of the exponent $b$\  
are given in table \ref{tab:fit_results}.

\begin{table}[h]
\caption{Results from the power law fits to the enhancement distributions (see text).  
\label{tab:fit_results}}
\begin{center}
\begin{tabular}{ccccccc}
\hline
 & $\Lambda$ & $\PgXm$ & $\PgOm$ & $\overline\Lambda$ & $\PagXp$ &  $\PagOp$ \\ \hline
 $b$ & $ 0.21\pm0.05$& $0.45\pm0.04$ & $0.45\pm0.28$ & $ 0.01\pm0.07$ & $0.47\pm0.10$ &   $0.49\pm0.29$ \\  
\hline
\end{tabular}
\end{center}
\end{table}
  

From figure \ref{enh_fit_few} 
and table \ref{tab:fit_results} one can see that for all multistrange hyperons and antihyperons 
we measure similar values of the exponent $b$. The $\Lambda$\ enhancement is slightly less centrality 
dependent, while the $\overline\Lambda$\ enhancement is compatible with being flat with centrality. 
Given the similarity of the behaviour of doubly and triply-strange hyperons and antihyperons, the difference 
between $\Lambda$ and $\overline\Lambda$ is surprising. 
This could be due to a difference in the initial production mechanism, or it could be the consequence 
of, e.g., a  centrality-dependent \PagL\ absorption in a nucleon-rich environment.
It is worth recalling, as described in section 4, that within the NA57 acceptance   
$\overline\Lambda$ is the only strange baryon for which we observe a significantly non-flat rapidity 
distribution, which we 
parametrised as a Gaussian~\cite{rapidity}. 

Extrapolating down to 1 the curves of figure  \ref{enh_fit_few}, all of them miss the pPb points by 
more than 6 standard deviations, with the exception of that of $\PagXp$ which is fully compatible with
the pPb point. They also miss the pBe points by a number of standard deviations ranging from 2 
to 27. As already mentioned, we observe scaling between pBe and pPb; therefore, 
our data disfavour the hypothesis 
of a regular increase (power law-type) of the enhancements with centrality starting from pA.

\section{Conclusions}

The hyperon yields per wounded nucleon at 158 $A$ GeV/$c$ are enhanced with respect to pBe 
and pPb interactions, these two being compatible with $N_{\rm wound}$ scaling. As already observed by 
the predecessor WA97 experiment, the global enhancement in the full centrality region (53\%  most 
central Pb--Pb events) increases with the content of valence strange quarks, reaching a factor of about 
20 for the triply-strange $\Omega$. This pattern is consistent with the QGP creation hypothesis, while it 
is hard to accomodate it into standard hadronic transport models. We observe a significant 
centrality dependence of the strangeness enhancement for all hyperons and anti-hyperons except for the $\overline\Lambda$. The  enhancements 
increase with centrality according to a power law whose exponents are compatible for 
multistrange hyperons and antihyperons, a factor two smaller for the $\Lambda$, 
while the $\overline\Lambda$ enhancement exhibits no centrality dependence. 
A saturation of the yields per participant 
cannot be ruled out,  
at least for the $\simeq10\%$ most central collisions.

%

\end{spacing}
\end{document}